\RequirePackage[l2tabu, orthodox]{nag}
\RequirePackage{snapshot}

\documentclass[9pt,twocolumn]{extarticle}

\makeatletter
\if@twocolumn
  \usepackage[dvips,letterpaper,top=0.5in, bottom=0.5in, left=0.75in, right=0.5in,includefoot,heightrounded]{geometry}
\else
  \usepackage[dvips,letterpaper,margin=1in,includefoot,heightrounded]{geometry}
\fi

\usepackage{srcltx}

\usepackage{amsmath}
\usepackage{amssymb,amsfonts}

\usepackage{abstract}

\usepackage{graphicx}
\usepackage{epsfig}
\usepackage[usenames,dvipsnames]{color}
\usepackage[usenames,dvipsnames]{xcolor}
\usepackage{subfigure}

\usepackage{booktabs}

\usepackage{setspace}
\usepackage{flushend}
\usepackage{multicol}

\usepackage{cite}
\usepackage{url}
\usepackage[normalem]{ulem}

\usepackage{enumerate}

\usepackage{hyperref}

\usepackage{multirow}

\usepackage[sc,tiny]{titlesec}

\usepackage{microtype}

\makeatletter

\newenvironment{rsmallmatrix}{\null\,\vcenter\bgroup
  \Let@\restore@math@cr\default@tag
  \baselineskip6\ex@ \lineskip1.5\ex@ \lineskiplimit\lineskip
  \ialign\bgroup\hfil$\m@th\scriptstyle##$&&\thickspace\hfil
  $\m@th\scriptstyle##$\crcr
}{%
  \crcr\egroup\egroup\,%
}
\makeatother

\title{%
Low-complexity Image and Video Coding Based on an Approximate Discrete Tchebichef Transform
}

\author{%
Paulo A. M. Oliveira%
\thanks{Paulo A. M. Oliveira
was with the
Signal Processing Group,
Departamento de Estat\'istica,
Universidade Federal de Pernambuco,
Recife, PE, Brazil;
Multimedia Communications and Signal Processing,
University of Erlangen--Nuremberg,
Erlangen, BY, Germany.}
\and
Renato~J.~Cintra%
\thanks{Renato~J.~Cintra
is with the
Signal Processing Group,
Universidade Federal de Pernambuco,
Caruaru, PE, Brazil.
He was with
\'Equipe Cairn,
INRIA-IRISA, Universit\'e de Rennes, Rennes, France;
and
LIRIS, Institut National des Sciences Appliqu\'ees,
Lyon, France
(e-mail: rjdsc@de.ufpe.br).}
\and
F\'abio~M.~Bayer%
\thanks{%
F\'abio~M.~Bayer
is with the
Departamento de Estat\'istica
and LACESM,
Universidade Federal de Santa Maria,
Santa Maria, RS, Brazil
(e-mail: bayer@ufsm.br).}
\and
Sunera~Kulasekera%
\thanks{Sunera Kulasekera and Arjuna Madanayake
were with the
Department of Electrical and Computer Engineering
at the University of Akron, OH.}
\and
Arjuna~Madanayake$^\S$
\and
V\'itor A. Coutinho
\thanks{V\'itor A. Coutinho was with
	the Signal Processing Group,
	Departamento de Estat\'{\i}stica, UFPE,
	Recife, Brazil.}
}

\date{}

\newcommand{\myabstract}{%
The usage of linear transformations has great relevance
for data decorrelation applications, like image and video compression.
In that sense,
the discrete Tchebichef transform (DTT)
possesses useful coding and decorrelation properties.
The DTT transform kernel does not depend on the input data
and fast algorithms can be developed to real time applications.
However,
the DTT fast algorithm presented in literature possess
high computational complexity.
In this work,
we introduce a new low-complexity approximation for the DTT.
The fast algorithm of the proposed transform is multiplication-free
and requires a reduced number of additions and bit-shifting operations.
Image and video compression simulations
in popular standards
shows good performance of the proposed transform.
Regarding hardware resource consumption for FPGA shows 43.1\%
reduction of configurable logic blocks
and ASIC place and route realization shows 57.7\% reduction in
the area-time figure
when compared with the \mbox{2-D} version of the exact DTT.
}

\newcommand{\mykeywords}{%
Approximate transforms,
discrete Tchebichef transform,
fast algorithms,
image and video coding
}

\begin{document}

\makeatletter
\if@twocolumn

\twocolumn[%
  \maketitle
  \begin{onecolabstract}
    \myabstract
  \end{onecolabstract}
  \begin{center}
    \small
    \textbf{Keywords}
    \\\medskip
    \mykeywords
  \end{center}
  \bigskip
]
\saythanks

\else

  \maketitle
  \begin{abstract}
    \myabstract
  \end{abstract}
  \begin{center}
    \small
    \textbf{Keywords}
    \\\medskip
    \mykeywords
  \end{center}
  \bigskip
  \onehalfspacing
\fi

\section{Introduction}

Discrete variable orthogonal polynomials
emerge as solutions of
several hypergeometric
difference equations~\cite{Nikiforov1991poly_discrete}.
Classic applications
of this class of orthogonal polynomials
include functional analysis~\cite{Dragnev1997func_analysis}
and graphs~\cite{Camara2009graphs}.
Additionally,
such polynomials
are employed
in the computation of moment functions~\cite{Zhu2010orth_moments},
which are largely used in
image processing~\cite{Goshtasby1985moment1,Heywood1995moment2,Markandey1992moment3}.
For instance,
the discrete Tchebichef moments~\cite{DTM},
which are derived from the discrete Tchebichef polynomials,
form a set of orthogonal moment functions.
Such functions are not
discrete approximation based on
continuous functions;
they are naturally
orthogonal over the discrete domain.

The Tchebichef moments have been used for
quantifying image block artifact~\cite{Leida2014artifact_tchebichef},
image recognition~\cite{Rose2009shape_tchebichef, Zhang2010img_recog, Li2011recognition_tchebichef},
blind integrity verification~\cite{Roux2012blind},
and
image compression~\cite{Swamy2008DTT, Swamy2013ITT, Mukundan2010img_compress, Li2012cs_tchebichef, Senapati2014listlessDTT}.
In the data compression context,
bi-dimensional (\mbox{2-D}) moments
are computed by means
of
the
\mbox{2-D}
discrete Tchebichef transform (DTT).
In fact,
the 8-point DTT
can achieve
better performance
when comparison with
the discrete cosine transform (DCT)~\cite{Ahmed1974DCT},
in terms of average bit length
as reported in~\cite{Ernawan2013quantization_tchebichef,Mukundan2010img_compress,Senapati2011DTT_coding}.
Moreover,
the 8-point DTT-based
embedded encoder proposed in~\cite{Senapati2014listlessDTT},
shows
improved image quality and
reduced encoding/decoding time
in comparison with state-of-the-art DCT-based embedded coders.
The 8-point DTT
has also been
employed in blind forensics,
as a tool to determine
the integrity of medical imagery subject
to filtering and compression~\cite{Roux2012blind}.

However,
the exact DTT possesses high arithmetic complexity,
due to its significant amount of additions
and float-point multiplications.
Such multiplications
are known to be more demanding computational
structures %
than additions or fixed-point multiplications,
both in \emph{software} and \emph{hardware}.
Thus,
the higher computational complexity
of the DTT
precludes its applications
in
low power consumption systems~\cite{Ernawan2011mobile_tchebichef, Li2008WSN}
and/or
real-time processing,
such as video streaming~\cite{Meng2005realtime_video,Friedman2013video_low_power}.
Therefore,
fast algorithms
for the DTT
could improve its computational efficiency.
A comprehensive literature search
reveals
only
two fast algorithms for
the 4-point DTT~\cite{Mukundan2007FDTT,Swamy2008DTT}
and
one for the 8-point DTT~\cite{Swamy2013ITT}.
Although
these fast algorithms possess
lower arithmetic complexities
when compared with the direct DTT calculation,
they still possess high arithmetic complexity,
requiring a significant amount of additions
and
bit-shifting operations.

In a comparable scenario,
the computation of DCT-based transforms---which
has been employed in
several popular coding schemes
such as
JPEG~\cite{Wallace1992JPEG},
\mbox{MPEG-2}~\cite{MPEG-2},
H.261~\cite{H.261},
H.263~\cite{H.263},
H.264~\cite{H264_book},
HEVC~\cite{Sullivan2012HEVC,Bossen2012HEVC_impementation},
and VP9~\cite{VP9}---has profited from
matrix approximation theory~\cite{Haweel2001SDCT, CB2011RDCT, CB2012MRDCT, cintra2014dct_aprox, BAS2008_EL, BAS2011, BAS2013}.
In this context,
discrete transforms are not exactly calculated,
but instead
an approximate, low-cost computation,
is performed.
The approximations are designed
in such a way to
allow similar spectral and coding characteristics
as well as lower arithmetic complexity.
Usually,
approximations are multiplierless,
requiring
only addition and bit-shifting operations
for its computation.
In~\cite{Oliveira2015Tchebichef},
a multiplierless approximation
for the 8-point DTT is proposed.
To the best of our knowledge,
this is the only DTT aproximation
archived in literature.

The aim of this work
is to introduce
an
efficient low-complexity approximation
for the 8-point DTT
capable of outperforming~\cite{Oliveira2015Tchebichef}.
To derive multiplierless approximate DTT matrix,
a multicriteria optimization problem
is sought,
combining
different coding metrics:
coding gain and transform efficiency.
Additionally,
a fast algorithm for
efficient
computation of the sought
approximation is also pursued.
For coding performance evaluation,
we propose two computational experiments:
(i)~a JPEG image compression simulation
and
(ii)~a video coding experiment
which consists of
embedding the sought approximation into the \mbox{H.264}/AVC standard.

The paper unfolds as follows.
Section~\ref{sec:DTT}
reviews the mathematical background of the DTT.
Section~\ref{sec:approx}
introduces
a parametrization of the DTT
to derive a family of DTT approximations
and
sets up
an
optimization problem
to identify optimal approximations.
In Section~\ref{sec:eval},
we assess the obtained approximation
in terms of coding performance,
proximity with the exact transform,
and
computation cost.
Moreover,
a fast algorithm for the proposed approximate DTT is introduced.
Section~\ref{sec:img_compress}
shows the results of the image and video compression simulations.
Section~\ref{section-hardware} shows hardware resource consumption comparison with the exact DTT for both FPGA and ASIC realizations.
A discussion and final remarks
are shown in Section~\ref{section-conclusion}.

\section{Discrete Tchebichef Transform}
\label{sec:DTT}

\subsection{Discrete Tchebichef Polynomials}

The discrete Tchebichef polynomials
are a set of discrete variable orthogonal polynomials~\cite{HTF53}.
The $k$th order discrete Tchebichef polynomials
are given by the following closed form
expression~\cite{Swamy2008DTT}:
\begin{align*}
t_k[n]
=
(1-N)_{k}
\cdot
{}_3 F_{2}(-k,-n,1+k;1,1-N;1)
,
\end{align*}
where
$n=0,1,\ldots,N-1$,
$_3 F_{2} (a_1, a_2, a_3; b_1, b_2;z)
= \sum_{n=0}^{\infty}
\frac{(a_1)_k (a_2)_k (a_3)_k}{(b_1)_k (b_2)_k}
\frac{z^k}{k!}$
is the generalized hypergeometric function
and
${(a)}_k = a(a+1)\cdots(a+k-1)$
is the descendant factorial.
Tchebichef polynomials
can be
obtained
according to the following
recursion~\cite{Swamy2008DTT}:
\begin{align*}
t_k[n]
=
&
\left[\frac{2k-1}{k}t_k[1]\right]t_{k-1}[n]
-
\left[\frac{k-1}{k}\left(N^2-(k-1)^2\right)\right]t_{k-2}[n]
,
\end{align*}
for
$t_0[n] = 1$
and
$t_1[n] = 2n-N+1$.
Indeed,
the set $\{t_k[n]\}$,
$k=0,1,\ldots, N-1$,
is an orthogonal basis in respect with the unit weight.
Consequently,
the discrete Tchebichef polynomials
satisfy the following
mathematical relation:
\begin{align*}
\sum_{i=0}^{N-1} t_i[n] t_j[n]
&=
\rho(j,N)
\cdot
\delta_{i,j}
,
\end{align*}
where
$\rho(k,N) =  \frac{(N+k)!}{(2k+1)\cdot(N-k-1)!}$
and
$\delta_{i,j}$ is the Kronecker delta function
which yields
$\delta_{i,j} = 1$,
if $i = j$,
and
$\delta_{i,j} = 0$,
otherwise.

\subsection{\mbox{2-D} Discrete Tchebichef Transform}

Let $\mathbf{f}[m,n]$,
$m,n = 0,1,\ldots,N-1$,
be
an intensity distribution from a discrete image
of size $N\times N$ pixels.
The \mbox{2-D} DTT
of $\mathbf{f}[m,n]$,
denoted by $\mathbf{M}[p,q]$,
$p,q = 0,1,\ldots,N-1$,
is given by~\cite{DTM,Swamy2008DTT}:
\begin{align}
\label{tmt2d}
\mathbf{M}[p,q]
=
\sum_{m,n=0}^{N-1}
\widetilde{t}_p[m]
\cdot
\widetilde{t}_q[n]
\cdot
\mathbf{f}[m,n]
,
\end{align}
where
$\widetilde{t}_k [n]$,
$k=0,1,\ldots,N-1$,
are the orthonormalized discrete Tchebichef polynomials
given by
$\widetilde{t}_k [n] = t_k[n]/\sqrt{\rho(k,N)}$.

Note that the transform kernel described in~\eqref{tmt2d}
is separable.
Hence,
the following is relation holds true:
\begin{align*}
\mathbf{M}[p,q]
=
\sum_{m=0}^{N-1}
\widetilde{t}_p[m]
\sum_{n=0}^{N-1}
\widetilde{t}_q[n]
\cdot
\mathbf{f}[m,n]
,
\end{align*}
for $p,q=0,1,\ldots,N-1$.
Therefore,
the transform-domain
coefficients of $\mathbf{f}$
can be calculated
by the following matrix operation:
\begin{align}
\label{direta}
\mathbf{M}
=
\mathbf{T}
\cdot
\mathbf{f}
\cdot
\mathbf{T}^\top
,
\end{align}
where
$\mathbf{T}$
is the $N$-point unidimensional DTT matrix
given by
\begin{align*}
\mathbf{T} =
\left[
\begin{smallmatrix}
\widetilde{t}_0[0] & \widetilde{t}_0[1] & \cdots & \widetilde{t}_0[N-1] \\
\widetilde{t}_1[0] & \widetilde{t}_1[1] & \cdots & \widetilde{t}_1[N-1] \\
\vdots  & \vdots  & \ddots & \vdots  \\
\widetilde{t}_{N-1}[0] & \widetilde{t}_{N-1}[1] & \cdots & \widetilde{t}_{N-1}[N-1]
\end{smallmatrix}
\right]
.
\end{align*}

The matrix operations induced by~\eqref{direta}
represents the \mbox{2-D} DTT.
Because of the kernel separation
property,
the~\mbox{2-D} DTT
can be calculated
by means the successive
applications
of the  \mbox{1-D} DTT
to
the rows of $\mathbf{f}$;
and then
to columns of the
resulting intermediate matrix.
The original intensity distribution $\mathbf{f}$
can be recovered by the inverse procedure:
\begin{align*}
\mathbf{f}
=
&
\mathbf{T}^{-1}
\cdot
\mathbf{M}
\cdot
(\mathbf{T}^{-1})^\top
=
\mathbf{T}^\top
\cdot
\mathbf{M}
\cdot
\mathbf{T}
.
\end{align*}
The last equality above stems
from the DTT orthogonality property:
$\mathbf{T}^{\top}=\mathbf{T}^{-1}$~\cite{Swamy2008DTT}.
Therefore,
the same structure can be used
at the forward transform as well in the inverse.

For $N=4$ and $N=8$,
we have the particular cases of interest
in the context of image and video coding.
Thus,
the 4- and 8-point DTT matrices are,
respectively,
furnished by:
\begin{align*}
\mathbf{T}_4
= &
\mathbf{F}_4
\cdot
\left[
\begin{rsmallmatrix}
  1 & 1 & 1 & \phantom{-}1 \\
	-3 & -1 & 1 & 3 \\
	1 & -1 & -1 & 1 \\
	-1 & 3 & -3 & 1 \\
\end{rsmallmatrix}
\right]
\end{align*}
and
\begin{align*}
\mathbf{T}_8
= &
\frac{1}{2}
\cdot
\mathbf{F}_8
\cdot
\left[
\begin{rsmallmatrix}
1 & 1 & 1 & 1 & 1 & 1 & 1 & \phantom{-}1 \\
-7 & -5 & -3 & -1 & 1 & 3 & 5 & 7 \\
7 & 1 & -3 & -5 & -5 & -3 & 1 & 7 \\
-7 & 5 & 7 & 3 & -3 & -7 & -5 & 7 \\
7 & -13 & -3 & 9 & 9 & -3 & -13 & 7 \\
-7 & 23 & -17 & -15 & 15 & 17 & -23 & 7 \\
1 & -5 & 9 & -5 & -5 & 9 & -5 & 1 \\
-1 & 7 & -21 & 35 & -35 & 21 & -7 & 1
\end{rsmallmatrix}
\right],
\end{align*}
where
$\mathbf{F}_4 =
\operatorname{diag}
\left(
\frac{1}{2},
\frac{1}{\sqrt{20}},
\frac{1}{2},
\frac{1}{\sqrt{20}}
\right)
$
and
$\mathbf{F}_8 =
\operatorname{diag}
\left(
\frac{1}{\sqrt{2}},
\frac{1}{\sqrt{42}},
\frac{1}{\sqrt{42}},
\frac{1}{\sqrt{66}},
\frac{1}{\sqrt{142}},
\frac{1}{\sqrt{546}},
\frac{1}{\sqrt{66}},
\frac{1}{\sqrt{858}}
\right)
$.
We observe that $\mathbf{T}_4$ and $\mathbf{T}_8$
are written as a result from the product of an integer matrix
and a diagonal matrix which requires float-point representation.

\section{DTT Approximations and Coding Optimality}
\label{sec:approx}

In this section,
we
aim at proposing
an extremely low-complexity
DTT approximation.
Our methodology
consists of
generating a class of
parametric
approximate matrices
and then
identify
the optimal
class member
in terms of coding performance.

\subsection{Related Work}

To the best of our knowledge,
the only DTT approximation
archived in literature
was proposed in~\cite{Oliveira2015Tchebichef}.
That approximation was obtained by
means of a parameterization of integer functions
combined
with
a normalization of transformation matrix columns.
The derived approximation in~\cite{Oliveira2015Tchebichef}
furnishes good coding capabilities,
but
it lacks orthogonality or near-orthogonality properties.
As a consequence,
the forward and inverse transformations
are quite distinct and possess unbalanced computational complexities.

\subsection{Parametric Low-complexity Matrices}

In~\cite{Haweel2001SDCT,CB2011RDCT,cintra2014dct_aprox,malvar2003H264matrix},
DCT approximations were proposed
according
to following operation:
\begin{align*}
\operatorname{int}(\alpha \cdot \mathbf{C}),
\end{align*}
where $\operatorname{int}(\cdot)$ is an integer function,
$\alpha$ is a real scaling factor,
and
$\mathbf{C}$ é is the exact DCT matrix.
Usual integer functions
include
the
floor,
ceiling,
signal,
and
rounding functions~\cite{cintra2014dct_aprox}.
In this work,
these functions operate
element-wise
when applied to a matrix argument.

A similar approach is sought for the proposed DTT approximation.
However,
in contrast with the DCT,
the rows of DTT matrix (basis vectors)
have a widely varying dynamic range.
Thus,
the integer function
may excessively
penalize the rows with small dynamic range.
To compensate this phenomenon,
we normalize the rows of~$\mathbf{T}_4$ and~$\mathbf{T}_8$
according
to left multiplications
by
$
\mathbf{D}_4
=
\operatorname{diag}
\left(
2,
\frac{\sqrt{20}}{3},
2,
\frac{\sqrt{20}}{3}
\right)$
and
$\mathbf{D}_8
=
\operatorname{diag}
\left(
\sqrt{8},
\frac{\sqrt{168}}{7},
\frac{\sqrt{168}}{7},
\frac{\sqrt{264}}{7},
\frac{\sqrt{568}}{13},
\frac{\sqrt{2184}}{23},
\frac{\sqrt{264}}{9},
\frac{\sqrt{3432}}{35}
\right)
$,
respectively.

The sought approximations
are required to possess extremely
low complexity.
One way of ensuring
this property
is to adopt an integer function
whose co-domain
is a set of low-complexity
integer.
In the DCT literature,
common
sets
are:
$\mathcal{P}_0=\{\pm1\}$~\cite{Haweel2001SDCT},
$\mathcal{P}_1=\{0,\pm1\}$~\cite{CB2011RDCT},
and
$\mathcal{P}_2=\{0,\pm1,\pm2\}$~\cite{cintra2014dct_aprox}.
Note that elements from these sets
have very simple realization in hardware;
implying multiplierless designs
with only addition
and bit-shifting operations~\cite{blahut_book}.

Adopting $\mathcal{P}_2$,
we
have that
a suitable integer function is given by:
\begin{align*}
\operatorname{round}
:
[-1,1]
&
\to
\mathcal{P}_2=\left\{0,\pm1,\pm2\right\},
\\
x
&
\mapsto
\operatorname{round}(\alpha\cdot x)
,
\qquad
0<\alpha < 5/2
,
\end{align*}
where
$
\operatorname{round}(x)
=
\operatorname{sign}(x)
\cdot
\lfloor
|x|
+
\frac{1}{2}
\rfloor
$
is the rounding function
as implemented
in
MATLAB~\cite{MATLAB},
Octave~\cite{octave},
and
Python~\cite{python}
programming languages.
Following the methodology described in~\cite{cintra2014dct_aprox},
we
obtain the following parametric
class of matrices:
\begin{align}
\label{equation-class}
\mathbf{T}_N(\alpha)
=
\operatorname{round}
(\alpha \cdot \mathbf{D}_N \cdot \mathbf{T}_N),
\quad
N\in\{4,8\}
.
\end{align}

\subsection{DTT Approximation}

A given low-complexity matrix
$\mathbf{T}_N(\alpha)$
can be used to approximate the DTT matrix
by means of
orthogonalization or quasi-orthogonalization
as
described in~\cite{CB2011RDCT,CB2012MRDCT,cintra2014dct_aprox}.
As a result,
an approximation for $\mathbf{T}_N$,
referred to as $\hat{\mathbf{T}}_N(\alpha)$,
can be obtained by:
\begin{align}
\label{equation-approximation}
\hat{\mathbf{T}}_N(\alpha)
=
\mathbf{S}_N(\alpha)
\cdot
\mathbf{T}_N(\alpha)
,
\end{align}
where
$
\mathbf{S}_N(\alpha)
=
\sqrt{
\left\{
\operatorname{ediag}
(
\mathbf{T}_N(\alpha)
\cdot
\mathbf{T}_N(\alpha)^\top
)
\right\}^{-1}
}
$
is a diagonal matrix,
$\operatorname{ediag}(\cdot)$
returns
a diagonal matrix
with
the diagonal entries of its argument
and
$\sqrt{\cdot}$
is the matrix
element-wise square root operator~\cite{cintra2014dct_aprox}.
If
\begin{align}
\label{equation-orthogonality-criterion}
\mathbf{T}_N(\alpha)
\cdot
\mathbf{T}_N(\alpha)^\top
=
[\text{diagonal matrix}]
\end{align}
holds true,
then $\hat{\mathbf{T}}_N(\alpha)$ is an orthogonal matrix~\cite{seber_matrix}.
Otherwise,
it is possibly a near orthogonal matrix~\cite{cintra2014dct_aprox}.
An approximation is said quasi-orthogonal
when the deviation from diagonality
of
$\mathbf{T}_N(\alpha)
\cdot
\mathbf{T}_N(\alpha)^\top$
is considered small.
Let $\mathbf{A}$ be a square real matrix.
The deviation from diagonality~$\delta(\mathbf{A})$
is given by~\cite{flury86}:
\begin{align*}
\delta(\mathbf{A})
=
1
-
\frac
{\|\operatorname{ediag}(\mathbf{A})\|_{F}}
{\|\mathbf{A}\|_{F}}
,
\end{align*}
where
$\|\cdot\|_{F}$
is the Frobenius norm for matrices~\cite{seber_matrix}.
In the context of image compression,
a deviation from diagonality value below
$
1
-
\frac{2}{\sqrt{5}}
\approx
0.1056
$
indicates quasi-orthogonality~\cite{cintra2014dct_aprox,Haweel2001SDCT}.

\subsection{Optimization Problem}

Now our goal is %
to identify
in the family~$\mathbf{T}_N(\alpha)$
the matrix that furnishes
the best approximation.
We adopted
two metrics as figures of merit
to guide the optimal choice:
(i)~the unified coding gain~$C_g$~\cite{Goyal2001coding,Katto1991cg}
and
(ii)~the transform efficiency~$\eta$~\cite{britanak_book}.
These metrics are relevant,
because they quantify the transform capacity of
removing
signal redundancy,
as well
as
data
compression and
decorrelation~\cite{britanak_book}.

Hence,
following the methodology in~\cite{tablada2015feig_winograd},
we propose the following
multicriteria optimization problem:
\begin{align*}
\alpha^\ast
=
\arg
\,
\max_{0<\alpha < 5/2}
\Big\{
C_g
\Big(
\hat{\mathbf{T}}_N(\alpha)
\Big),
\eta
\Big(
\hat{\mathbf{T}}_N(\alpha)
\Big)
\Big\}
,
\quad
N\in\{4,8\}
,
\end{align*}
where
$\alpha^\ast$
is the scaling parameter
that
results
in the optimal low complexity
matrix
$\mathbf{T}_N^\ast \triangleq \hat{\mathbf{T}}_N(\alpha^\ast)$
according to~\eqref{equation-class}.

The above optimization problem
is not analytically tractable.
Thus,
we resort to
exhaustive numerical search
to obtain~$\alpha^\ast$.
We consider linearly spaced values
of~$\alpha$
with a step of $10^{-3}$
in the interval~$0<\alpha < 5/2$.
For $N=4$
and $N=8$,
we obtain that
optimality is found
in the intervals
$(\frac{3}{2}, \frac{5}{2})$
and
$(\frac{23}{14}, \frac{69}{34})$,
respectively.
Therefore,
any value of~$\alpha$ in the aforementioned
intervals effects the same approximations.
For operational reasons,
we selected
$\alpha = 2$.
Thus,
the resulting
low-complexity matrices are given below:
\begin{align*}
\mathbf{T}_4^\ast
=
\left[
\begin{rsmallmatrix} 1 & 1 & 1 & \phantom{-}1 \\
       -2 & -1 & 1 & 2 \\
        1 & -1 & -1 & 1 \\
       -1 & 2 & -2 & 1 \\
\end{rsmallmatrix}
\right]
\quad
\text{and}
\quad
\mathbf{T}_8^\ast
=
\left[
\begin{rsmallmatrix}
 1 & 1 & 1 & 1 & 1 & 1 & 1 & \phantom{-}1 \\
       -2 & -1 & -1 & 0 & 0 & 1 & 1 & 2 \\
        2 & 0 & -1 & -1 & -1 & -1 & 0 & 2 \\
       -2 & 1 & 2 & 1 & -1 & -2 & -1 & 2 \\
        1 & -2 & 0 & 1 & 1 & 0 & -2 & 1 \\
       -1 &  2 & -1 & -1 & 1 & 1 & -2 & 1 \\
        0 & -1 & 2 & -1 & -1 & 2 & -1 & 0 \\
        0 & 0 & -1 & 2 & -2 & 1 & 0 & 0 \\
\end{rsmallmatrix}
\right]
.
\end{align*}
The associate
optimal approximations
are denoted by
$\hat{\mathbf{T}}_N^\ast \triangleq \hat{\mathbf{T}}_N(\alpha^\ast)$
and
can be computed
according to~\eqref{equation-approximation}.
Hence,
we obtain:
$
\hat{\mathbf{T}}_4^\ast
=
\operatorname{diag}
\left(
\frac{1}{2},
\frac{1}{\sqrt{10}},
\frac{1}{2},
\frac{1}{\sqrt{10}}
\right)
\cdot
\mathbf{T}_4^\ast
$
and
$
\hat{\mathbf{T}}_8^\ast
=
\operatorname{diag}
\left(
\frac{1}{\sqrt{8}},
\frac{1}{\sqrt{12}},
\frac{1}{\sqrt{12}},
\frac{1}{\sqrt{20}},
\frac{1}{\sqrt{12}},
\frac{1}{\sqrt{14}},
\frac{1}{\sqrt{12}},
\frac{1}{\sqrt{10}}
\right)
\cdot
\mathbf{T}_8^\ast
$

\section{Evaluation and Computational Complexity}
\label{sec:eval}

\subsection{Discussion}

The obtained
matrix $\mathbf{T}_4^\ast$
satisfies~\eqref{equation-orthogonality-criterion}
and
therefore $\hat{\mathbf{T}}_4^\ast$ is orthogonal.
In fact,
the proposed matrix
is identical to the
4-point integer transform for H.264 encoding
introduced by Malvar~\emph{et~al.}~\cite{malvar2003H264matrix}.
Therefore,
it is also an optimal approximate DTT.
Because the 4-point DCT approximation matrix by Malvar~\emph{et~al.}
was
submitted to in-depth analyses
in the context of video coding~\cite{H264_book},
such results also apply to~$\mathbf{T}_4^\ast$.
Therefore,
hereafter
we focus the forthcoming discussions
to the proposed 8-point
approximation~$\mathbf{T}_8^\ast$.

\subsection{Orthogonality and Invertibility}

The matrix~$\mathbf{T}_8^\ast$
does not satisfy~\eqref{equation-orthogonality-criterion}.
Therefore,
the associate approximation $\hat{\mathbf{T}}_8^\ast$
is not orthogonal,
i.e.
$
(\hat{\mathbf{T}}_8^\ast)^{-1}
\not=
(\hat{\mathbf{T}}_8^\ast)^\top
$.
As a consequence,
the inverse
does not
inherit
the low-complexity properties of
$\mathbf{T}_8^\ast$.
In other words,
the entries
of
$
(\mathbf{T}_8^\ast)^{-1}
$
are not in~$\mathcal{P}_2$.
Hence,
the proposed transform possesses
asymmetrical computational costs
when comparing
the direct and inverse operations~\cite{Cintra2002RHT}.
The inverse transformation
is a much required tool,
specially
for reconstruction
encoded images
back to the spatial domain~\cite{Wallace1992JPEG,Pennebaker1993JPEG}.

However,
$\mathbf{T}_8^\ast \cdot (\mathbf{T}_8^\ast)^\top$
has a low deviation from diagonality~\cite{flury86,cintra2014dct_aprox}:
only $0.024$---roughly
4.4 times less than the deviation
implied by the SDCT~\cite{Haweel2001SDCT},
which is taken
as the standard reference.
Therefore,
the following approximation
is valid:
$
(\hat{\mathbf{T}}_8^\ast)^{-1}
\approx
(\mathbf{T}_8^\ast)^\top
\cdot
\operatorname{diag}
\left(
\frac{1}{\sqrt{8}},
\frac{1}{\sqrt{12}},
\frac{1}{\sqrt{12}},
\frac{1}{\sqrt{20}},
\frac{1}{\sqrt{12}},
\frac{1}{\sqrt{14}},
\frac{1}{\sqrt{12}},
\frac{1}{\sqrt{10}}
\right)
$.
Moreover,
since diagonal matrices can be absorbed
into other computational steps~\cite{BAS2011,cintra2014dct_aprox,Oliveira2015Tchebichef},
$(\mathbf{T}_8^\ast)^{-1}$
can be
replaced
with
the low-complexity matrix
$(\mathbf{T}_8^\ast)^\top$.
This
has the advantage
of using
the same algorithm
for both forward
and inverse approximations.

\subsection{Performance Assessment}

The proposed approximations were
compared with their corresponding exact DTT
in terms of
coding performance
as measured according to
the coding gain
and
the transform efficiency.
For $N=8$,
we also include in our comparisons
the DTT approximation proposed in~\cite{Oliveira2015Tchebichef}.
The coding performance of
the proposed approximations
was evaluated
according to
the figures of merit $C_g$ and $\eta$.
Table~\ref{table:performance_codif}
displays the results.
The proposed approximations
are capable of furnishing
coding measures
very close to the exact transformations.
For comparison purposes,
the exact DCT
has its coding gain and transform efficiency of
8.83~dB
and
93.99,
respectively.

\begin{table*}
	\centering

	\caption{Performance assessment}
	\label{table:performance_codif}

	\begin{tabular}{llcccccc}
	\toprule
	$N$ & Method & $C_g$ (dB) & $\eta$ & MSE & $\epsilon$ & $d$ & $\delta$\\
	\midrule
	\multirow{2}{*}{$4$}
	& Exact DTT~\cite{Mukundan2007FDTT}  & 7.55 & 97.25 & - & - & - & 0\\
	& Proposed & 7.55 & 97.33 & 0.001 & 0.13 & 0.29\% & 0\\
	\midrule
	\multirow{3}{*}{$8$}
	& Exact DTT~\cite{Swamy2013ITT}  & 8.68 & 92.86 & - & - & - & 0\\
	& DTT Approx.~\cite{Oliveira2015Tchebichef}  & 5.51 & 83.51 & 0.015 & 3.32 & 12.61\% & 0.09\\
	& Proposed & 9.25 & 92.71 & 0.002 & 0.77 & 3.03\% & 0.024\\
	\bottomrule
	\end{tabular}

\end{table*}

Although our goal
is to derive good approximations
for coding,
we also analyzed the resulting approximations
in terms of proximity metrics.
We separated
the mean square error (MSE)~\cite{bovik2009mse},
the total energy error $\epsilon$~\cite{CB2011RDCT},
and
the transform distortion $d$~\cite{Chi2012distortion}.
All these measures
aim at
quantifying
the distance between the exact transformations
and their respective approximations.
Analytic expressions
for the MSE and $\epsilon$ are detailed in~\cite{tablada2015feig_winograd}.
We also evaluated the proximity of the proposed approximations
with respect to the exact DTT
according to the transform distortion measure suggested in~\cite{Chi2012distortion}.
This metric was originally proposed as the DCT distortion
in the context of DCT approximations
and
quantifies in percentage
a distance between
exact and approximate DCT.
Adapting it for the DTT,
we obtain the transform distortion
as follows:
\begin{align*}
d
(\hat{\mathbf{T}}_N^\ast)
=
\left\{
1
-
\frac{1}{N}
\cdot
\Big\|
\operatorname{ediag}
\Big\{
\mathbf{T}_N
\cdot
(\hat{\mathbf{T}}_N^\ast)^\top
\Big\}
\Big\|^{2}_{2}
\right\}
\times
100\%
,
\end{align*}
where
$\|\cdot\|_{2}$
is the euclidean norm for matrices~\cite{seber_matrix}.
Low values of distortion indicates proximity with the DTT.
As a comparison for $N=4$,
the 4-point DCT approximation proposed in~\cite{Bayer2013multless4point}
has a distortion of
$7.32\%$.
Proximity results are
also
shown in Table~\ref{table:performance_codif}.

\subsection{Fast Algorithm and Arithmetic Complexity}

Now we aim at deriving fast algorithms
for the obtained DTT approximations.
Being identical to the H.264 4-point DCT approximation,
the derived matrix~$\mathbf{T}_4^\ast$
was given fast algorithms in~\cite{malvar2003H264matrix}.
Although~$\mathbf{T}_8^\ast$
is multiplication free,
without a fast algorithm,
its direct implementation
requires
44~additions and
24~bit-shifting operations.
Thus we focus our efforts on the efficient computation
of $\mathbf{T}_8^\ast$.
To such end,
a sparse matrix factorization is sought,
where
the number of additions and bit-shifting operations
can be significantly reduced~\cite{blahut_book,oppenheim_book}.

The sparse matrix factorization
proposed in the manuscript
was derived from scratch
based on usual butterfly structures~\cite{blahut_book}.
We obtained the following decomposition:
\begin{align*}
\mathbf{T}^\ast_8
=
\mathbf{P}
\cdot
\mathbf{A}_2
\cdot
\mathbf{A}_1
\cdot
\mathbf{B}_8
,
\end{align*}
where
\begin{align*}
\mathbf{B}_8
&
=
\left[
\begin{rsmallmatrix}
1 & \phantom{-}0 & \phantom{-}0 & \phantom{-}0 & \phantom{-}0 & 0 & 0 & 1 \\
0 & 1 & 0 & 0 & 0 & 0 & 1 & 0 \\
0 & 0 & 1 & 0 & 0 & 1 & 0 & 0 \\
0 & 0 & 0 & 1 & 1 & 0 & 0 & 0 \\
0 & 0 & 0 & 1 & -1 & 0 & 0 & 0 \\
0 & 0 & 1 & 0 & 0 & -1 & 0 & 0 \\
0 & 1 & 0 & 0 & 0 & 0 & -1 & 0 \\
1 & 0 & 0 & 0 & 0 & 0 & 0 & -1 \\
\end{rsmallmatrix}
\right]
,
\\
\mathbf{A}_1
&=
\left[
\begin{rsmallmatrix}
0 & 0 & 1 & 0 & \phantom{-}0 & \phantom{-}0 & \phantom{-}0 & \phantom{-}0 \\
1 & 0 & 0 & 1 & 0 & 0 & 0 & 0 \\
0 & 1 & 0 & 0 & 0 & 0 & 0 & 0 \\
0 & -1 & 2 & -1 & 0 & 0 & 0 & 0 \\
2 & 0 & -1 & -1 & 0 & 0 & 0 & 0 \\
0 & 0 & 0 & 0 & 2 & -1 & 0 & 0 \\
0 & 0 & 0 & 0 & 1 & 1 & 0 & 0 \\
0 & 0 & 0 & 0 & 0 & 1 & 1 & 0 \\
0 & 0 & 0 & 0 & 0 & 0 & 2 & -1 \\
0 & 0 & 0 & 0 & 0 & 0 & 0 & -2 \\
\end{rsmallmatrix}
\right]
\\
\mathbf{A}_2
&
=
\left[
\begin{rsmallmatrix}
1 & \phantom{-}1 & 1 & \phantom{-}0 & \phantom{-}0 & \phantom{-}0 & 0 & 0 & \phantom{-}0 & \phantom{-}0 \\
0 & 1 & -2 & 0 & 0 & 0 & 0 & 0 & 0 & 0 \\
0 & 0 & 0 & 1 & 0 & 0 & 0 & 0 & 0 & 0 \\
0 & 0 & 0 & 0 & 1 & 0 & 0 & 0 & 0 & 0 \\
0 & 0 & 0 & 0 & 0 & 1 & 0 & 0 & 0 & 0 \\
0 & 0 & 0 & 0 & 0 & 0 & 1 & 1 & 0 & 1 \\
0 & 0 & 0 & 0 & 0 & 0 & 0 & -1 & 0 & 1 \\
0 & 0 & 0 & 0 & 0 & 0 & -1 & 0 & 1 & 0 \\
\end{rsmallmatrix}
\right]
,
\\
\mathbf{P}
&
=
\left[
\begin{rsmallmatrix}
1 & 0 & 0 & 0 & 0 & 0 & 0 & 0 \\
0 & 0 & 0 & 0 & 0 & 0 & 1 & 0 \\
0 & 0 & 0 & 1 & 0 & 0 & 0 & 0 \\
0 & 0 & 0 & 0 & 0 & 1 & 0 & 0 \\
0 & 1 & 0 & 0 & 0 & 0 & 0 & 0 \\
0 & 0 & 0 & 0 & 0 & 0 & 0 & 1 \\
0 & 0 & 1 & 0 & 0 & 0 & 0 & 0 \\
0 & 0 & 0 & 0 & 1 & 0 & 0 & 0 \\
\end{rsmallmatrix}
\right]
.
\end{align*}
Matrix $\mathbf{B}_8$
represents a layer of butterfly structures,
$\mathbf{A}_1$
and
$\mathbf{A}_2$
denote additive matrices with bit-shifting operations,
and
$\mathbf{P}$
represents
a final permutation,
which is cost-free.
The resulting
algorithm preserves
all algebraic and coding properties
of the direct computation,
while
requiring less arithmetic operations.
Moreover,
the factor of~2
of the first matrix row
can be absorbed into
the diagonal matrix.
The obtained factorization is represented
by the signal flow graph
shown in Figure~\ref{dfs}.
Such algorithm
reduces the
arithmetic cost of the proposed approximation
to
only
24~additions and six bit-shifting operations.

\begin{figure*}
\centering
\includegraphics{diagram_T.eps}
\caption{
Signal flow graph for $\mathbf{T}_8^\ast$.
Input data $x_n$, $n=0,1,\ldots,7$,
relates to output $X_m$, $m=0,1,\ldots,7$.
Dashed arrows and black nodes
represent multiplications by $-1$ and $2$,
respectively.}
\label{dfs}
\end{figure*}

Table~\ref{table:performance_compl}
compares the arithmetic complexity
of the discussed methods
evaluated
according to both
their respective
fast algorithms.
The additive and total arithmetic complexity
of the proposed approximation
are
45.5\% and 58.9\%
lower than the exact transform,
respectively.
Although,
the computational cost
of the proposed approximation
is slightly higher than the \emph{forward} DTT approximation
in~\cite{Oliveira2015Tchebichef},
it is important to notice that the inverse transformation
in~\cite{Oliveira2015Tchebichef}
is relatively more complex.
Considering the combination of forward and inverse transformation,
the design
in~\cite{Oliveira2015Tchebichef}
requires 49~additions,
whereas the proposed design
requires 48~additions.
Bit-shifting costs are virtually null,
because in a hardware implementation
they represent only wiring.
For comparison,
the popular
Loeffler DCT algorithm~\cite{Loeffler1989}
requires
11~floating-point multiplications
and
29~additions.
Although the actual approximation
consists of the multiplication of the low-complexity matrix
and
a diagonal matrix as shown in~\eqref{equation-approximation},
the multiplications introduced
by diagonal matrices
represent no
additional arithmetic complexity in image compression applications.
This is because
they can
be absorbed into the image quantization step~\cite{Wallace1992JPEG,Pennebaker1993JPEG}
of JPEG-like
image compression~\cite{CB2011RDCT,CB2012MRDCT,cintra2014dct_aprox,BAS2008_EL,BAS2011,BAS2013}.
Furthermore,
the new approximation is capable of
a better
coding performance and possesses
one order of magnitude lower proximity measure
as shown in Table~\ref{table:performance_codif}.
Therefore,
both performance
and
arithmetic complexity measures
are favorable to the
proposed approximation.

\begin{table*}
	\centering
	\caption{Fast algorithm arithmetic complexity comparison}
	\label{table:performance_compl}

  \begin{tabular}{lcccc}
    \toprule
      Method & Mult. & Adit. & Shifts & Total \\
    \midrule
     Exact DTT~\cite{Swamy2013ITT} & 0 & 44 & 29 & 73 \\
      Forward DTT approx.~\cite{Oliveira2015Tchebichef} & 0 & 20 & 0 & 20 \\
      Inverse DTT approx.~\cite{Oliveira2015Tchebichef} & 0 & 29 & 8 & 37 \\
      Proposed & 0 & 24 & 6 & 30 \\
	  \bottomrule
	\end{tabular}

\end{table*}

\section{Image and Video Compression Experiments}
\label{sec:img_compress}

In this section,
we perform two computation experiments.
The first one consists
of still image compression
considering a JPEG procedure.
The second simulation
assess the effectiveness
of the proposed approximations
under realistic video encoding
conditions.

\subsection{Image Compression}

We adopted the image compression
method
according to the JPEG standard~\cite{Wallace1992JPEG,Pennebaker1993JPEG}.
A set of 45
512$\times$512 8-bit images
were obtained
from a public image bank~\cite{imagens}
and submitted to processing.
The selected images
encompass a wide range of categories,
including
13~textures,
12~satellite images,
three human faces
and
several other miscellaneous scenarios.
For each image,
the luminance component
was extracted
and
subdivided
into 8$\times$8 blocks,
$\mathbf{I}_{i,j}$,
$i, j = 1, 2, \ldots, 64$.
After preprocessing,
each block
was submitted to
the the following
operation:
\begin{align*}
\mathbf{M}_{i,j}
=
\mathbf{P}
\cdot
\mathbf{I}_{i,j}
\cdot
\mathbf{P}^\top,
\end{align*}
where $\mathbf{M}_{i,j}$
is the \mbox{2-D}
transform-domain
data
and
$\mathbf{P}$
is a given
\mbox{1-D} transformation matrix,
such as the
exact or approximate DTT.
Then each subblock
$\mathbf{M}_{i,j}$
was element-wise divided
by
a quantization matrix,
yielding
the quantized JPEG coefficients
$\mathbf{J}_{i,j}$,
as follows:
\begin{align*}
\mathbf{J}_{i,j}
=
\operatorname{round}
\left(
\mathbf{M}_{i,j}
\oslash
\mathbf{Q}
\right)
,
\end{align*}
where
$
\mathbf{Q}
=
\lfloor
(
{S \cdot \mathbf{Q}_0 + 50}
)
/
{100}
\rfloor
$
is the quantization matrix,
$\lfloor \cdot \rfloor$ denotes the floor function,
the default quantization table
is
\begin{align*}
\mathbf{Q}_0
&=
\left[
\begin{rsmallmatrix}
16 & 11 & 10 & 16 & 24  & 40  & 51  & 61 \\
12 & 12 & 14 & 19 & 26  & 58  & 60  & 55 \\
14 & 13 & 16 & 24 & 40  & 57  & 69  & 56 \\
14 & 17 & 22 & 29 & 51  & 84  & 80  & 62 \\
18 & 22 & 37 & 56 & 68  & 109 & 103 & 77 \\
24 & 35 & 55 & 64 & 81  & 104 & 113 & 92 \\
49 & 64 & 78 & 87 & 103 & 121 & 120 & 101 \\
72 & 92 & 95 & 98 & 112 & 100 & 103 & 99
\end{rsmallmatrix}
\right]
,
\end{align*}
$S = 5000/QF$, if $QF < 50$,
and $200 - 2 \cdot QF$ otherwise,
and
$QF$
is the quality factor~\cite{Pennebaker1993JPEG}.
If $QF = 50$, then $\mathbf{Q} = \mathbf{Q}_0$.
Decreasing values of $QF$
lead to higher compression ratios
(with image total destruction at $QF=0$);
whereas
increasing values leads to lower compression ratios
(with best possible quality at $QF=100$).
In our experiments,
we adopted
$QF$
varying from 10 to 90
in steps of~5.
In the JPEG decoder,
each sub-block is initially
arithmetic decoded and dequantized
according to:
$
\hat{\mathbf{M}}_{i,j}
=
\mathbf{J}_{i,j}
\odot
\mathbf{Q}
$.
Then,
the sub-blocks are
inverse
transformed:
$
\mathbf{\hat{I}}_{i,j}
=
\mathbf{P}^{-1}
\cdot
\hat{\mathbf{M}}_{i,j}
\cdot
\mathbf{P}^{-\top}
$.

\begin{figure*}
\centering
\subfigure[]{\includegraphics[width=0.4\textwidth]{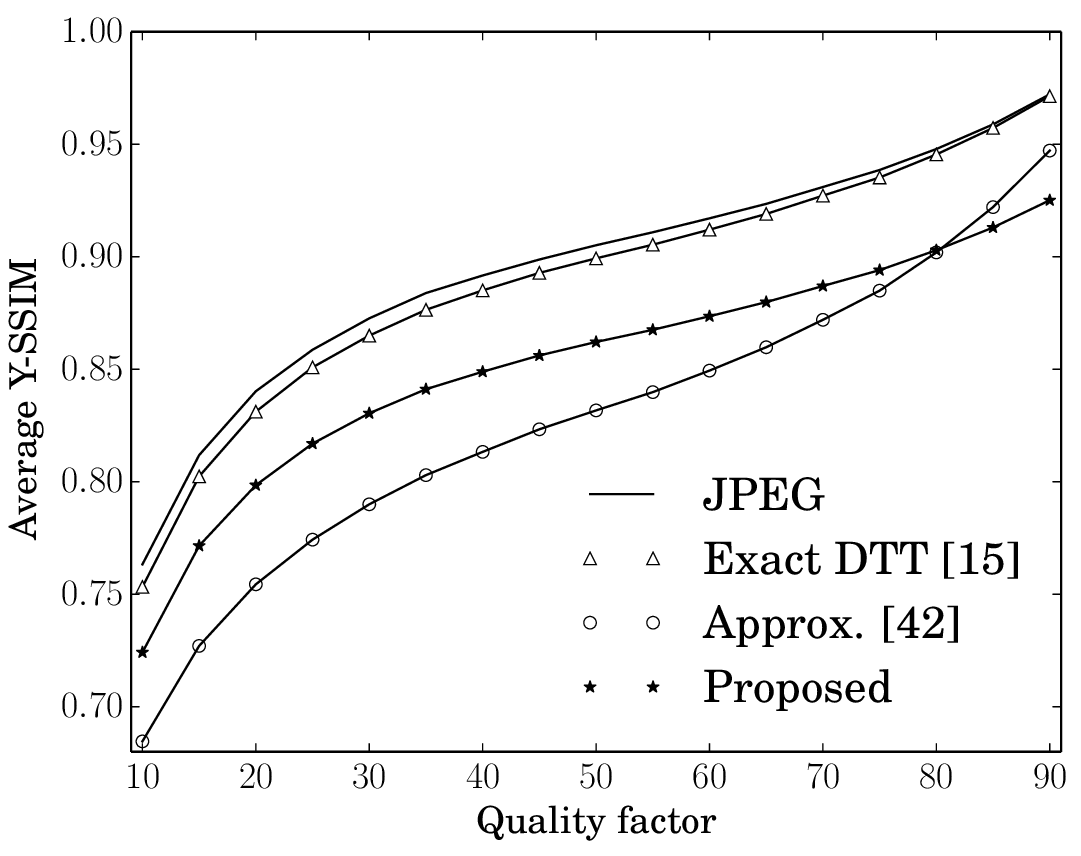}}
\subfigure[]{\includegraphics[width=0.4\textwidth]{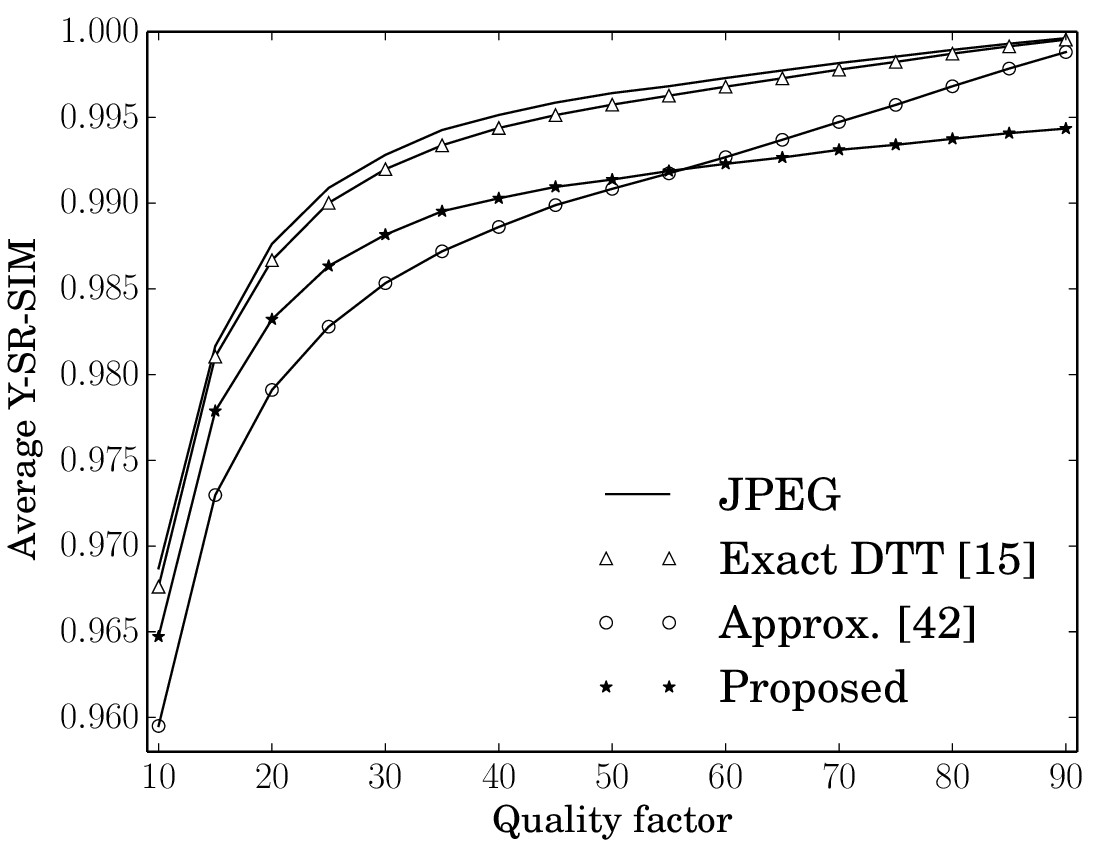}}
\caption{Average SSIM (top) and SR-SIM (bottom) measurements
for image compression
for the
considered
transforms
at several values of~$QF$.
}
\label{figure:qualidade}
\end{figure*}

Original and compressed images
were compared
for image degradation.
The structural similarity index (SSIM)~\cite{Bovik2004SSIM}
and the spectral residual based similarity (SR-SIM)~\cite{zhang2012SR_SIM}
were
separated
as image quality measures.
The SSIM takes into account
luminance, contrast, and the image structure
to quantify the image degradation, being consistent with
subjective quality measurements~\cite{Bovik2011SSIM}.
On its turn,
the SR-SIM
is based on the hypothesis that
the visual saliency maps of natural images
are closely related to their perceived quality.
This measure could outperform
several
state-of-the-art figures of merit
in experiments with standardized datasets~\cite{zhang2012SR_SIM}.
We did
not consider the peak signal-to-noise ratio (PSNR) as a quality measure,
because it is not a suitable metric
to capture the human perception
of image fidelity and quality~\cite{bovik2009mse}.
For each value of $QF$,
we considered
average measure values
instead of values from
particular images.
This approach is less prone to
variance effects and fortuitous data~\cite{kay_book,CB2011RDCT}.
For direct comparison,
we selected
the exact DTT~\cite{Swamy2013ITT},
the approximation proposed in~\cite{Oliveira2015Tchebichef},
and
the proposed approximation.
As an extra reference,
we also included
the results from the standard JPEG,
which is based on the exact DCT.
Figure~\ref{figure:qualidade}
displays the results.
For both selected measures,
the proposed approximation
performed
very closely
to the exact DTT,
specially at high compression ratios.
It could outperform
the DTT approximation in~\cite{Oliveira2015Tchebichef}
in terms of SSIM and SR-SIM
for
$QF<80$
and
$QF<55$,
respectively.
It shows that the proposed approximation is more efficient
in the scenario of high and moderate compression,
which are the very common cases~\cite{Pandit2013quality},
suitable for
low-power devices.
The approximation in~\cite{Oliveira2015Tchebichef}
could attain
a better performance
at low compression ratios
because it satisfies the perfect reconstruction property---albeit
at the expense of an inverse transformation
with higher arithmetic complexity.
On the other hand,
the proposed approximation explores the near-orthogonality property
which could excel in moderate to high compression scenarios---which are
often more relevant~\cite{Pandit2013quality}.

\begin{figure*}

\centering
\subfigure[Exact DTT~\cite{Swamy2013ITT},~$QF=15$]{\includegraphics[width=0.24\textwidth]{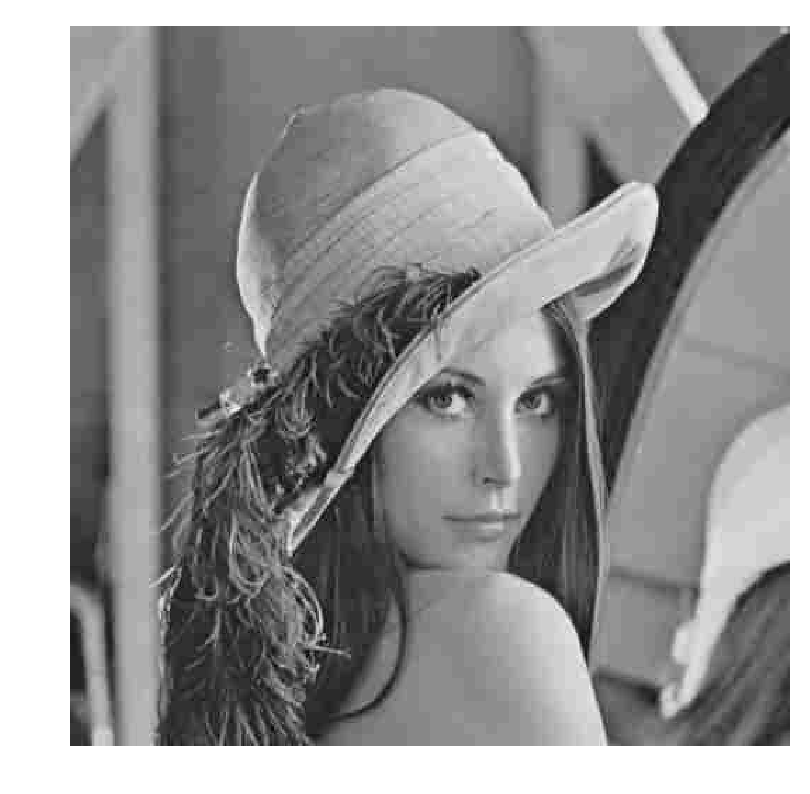}}
\subfigure[Exact DTT~\cite{Swamy2013ITT},~$QF=50$]{\includegraphics[width=0.24\textwidth]{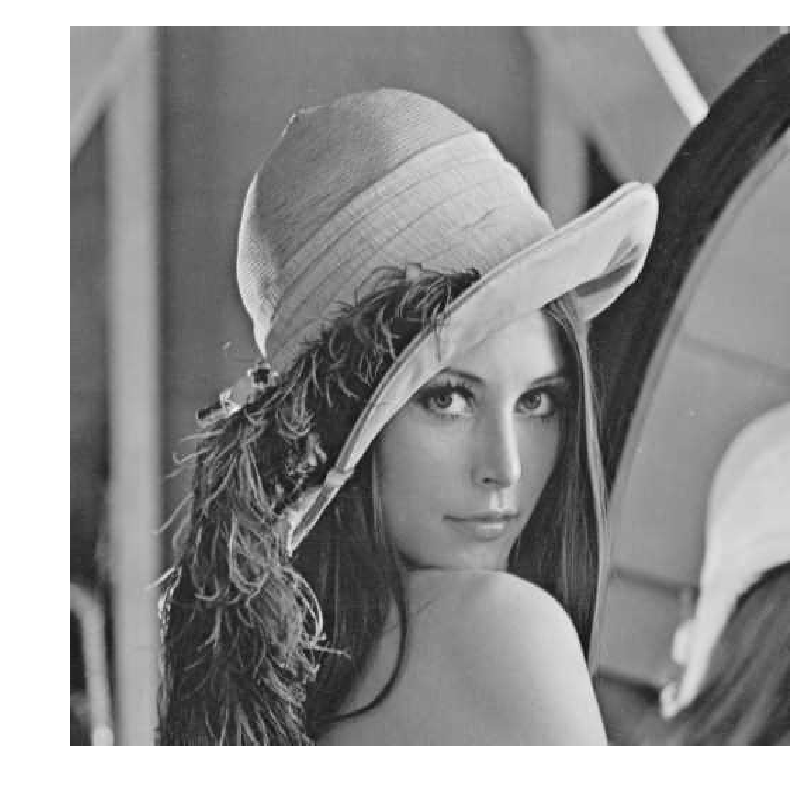}}
\subfigure[Proposed,~$QF=15$]{\includegraphics[width=0.24\textwidth]{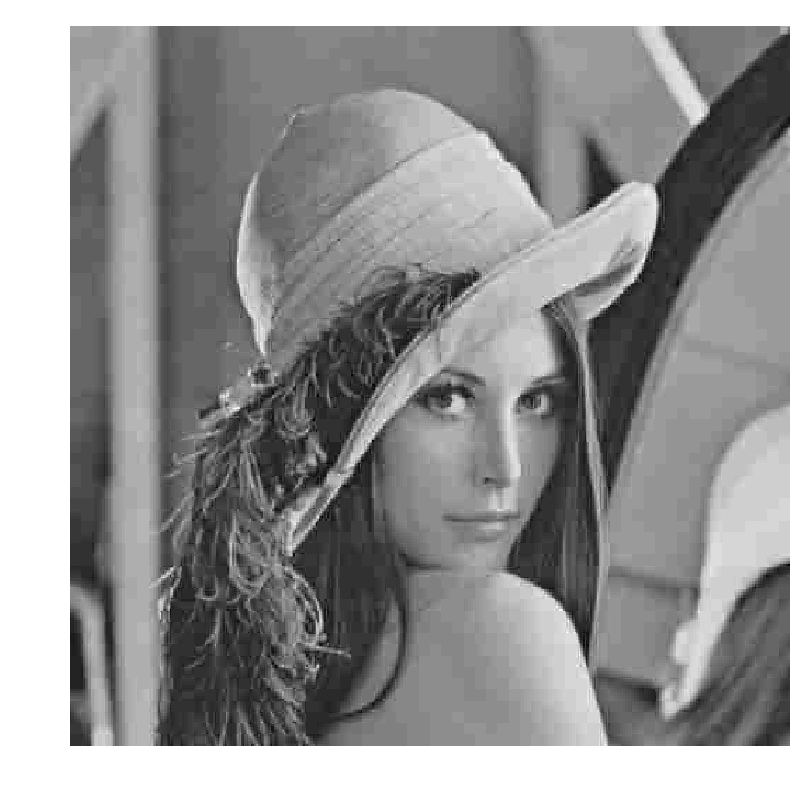}}
\subfigure[Proposed,~$QF=50$]{\includegraphics[width=0.24\textwidth]{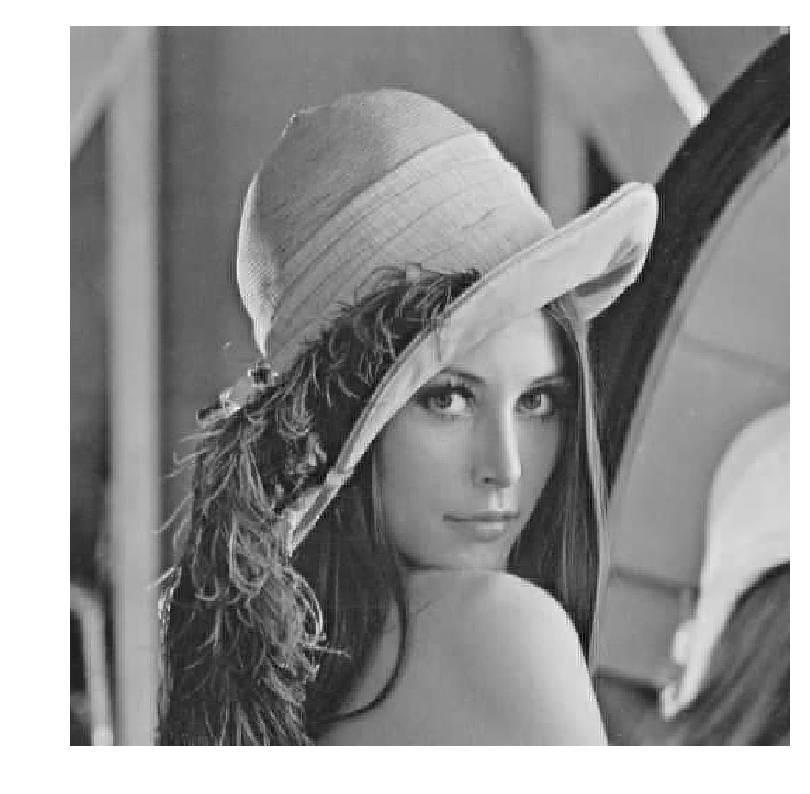}}
\caption{Compressed `Lena' image
for $QF = 15$ and $QF = 50$.}
\label{figure:lena}
\end{figure*}

For qualitative evaluation purposes,
Figure~\ref{figure:lena}
shows the compressed Lena image
according to the exact DTT
and the proposed approximation.
We adopted the scenario of high/moderate compression
with $QF=15$
and
$QF=50$,
respectively.
In both cases,
the approximate transform
was capable of producing
comparable results to the exact DTT with
visually similar images.

\subsection{Video Compression Simulation}
\label{sec:video_compress}

To evaluate the proposed transform performance in video coding,
we have embedded the DTT approximation
in the widely used
x264 software library~\cite{x264}
for encoding video streams into the \mbox{H.264}/AVC standard~\cite{H264_book}.
The default
8-point transform employed in \mbox{H.264} is
the following integer DCT approximation~\cite{H264_8transform}:
\begin{align*}
\mathbf{\hat{C}}
=
\frac{1}{8}
\cdot
\left[
\begin{rsmallmatrix} 8 & 8 & 8 & 8 & 8 & 8 & 8 & 8 \\
       12 & 10 & 6 & 3 & -3 & -6 & -10 & -12 \\
        8 & 4 & -4 & -8 & -8 & -4 & 4 & 8 \\
       10 & -3 & -12 & -6 & 6 & 12 & 3 & -10 \\
        8 & -8 & -8 & 8 & 8 & -8 & -8 & 8 \\
        6 & -12 & 3 & 10 & -10 & -3 & 12 & -6 \\
        4 & -8 & 8 & -4 & -4 & 8 & -8 & 4 \\
        3 & -6 & 10 & -12 & 12 & -10 & 6 & -3 \\
\end{rsmallmatrix}
\right].
\end{align*}
The
fast algorithm for the
above transformation
requires 32~additions and 14~bit-shifting
operations~\cite{H264_8transform}.
Therefore,
the proposed 8-point transform
requires 25\% less additions
and 57\% less bit-shifting operations
than the fast algorithm for $\mathbf{\hat{C}}$.

\begin{figure*}
\centering
\subfigure[]{\includegraphics[width=0.4\textwidth]{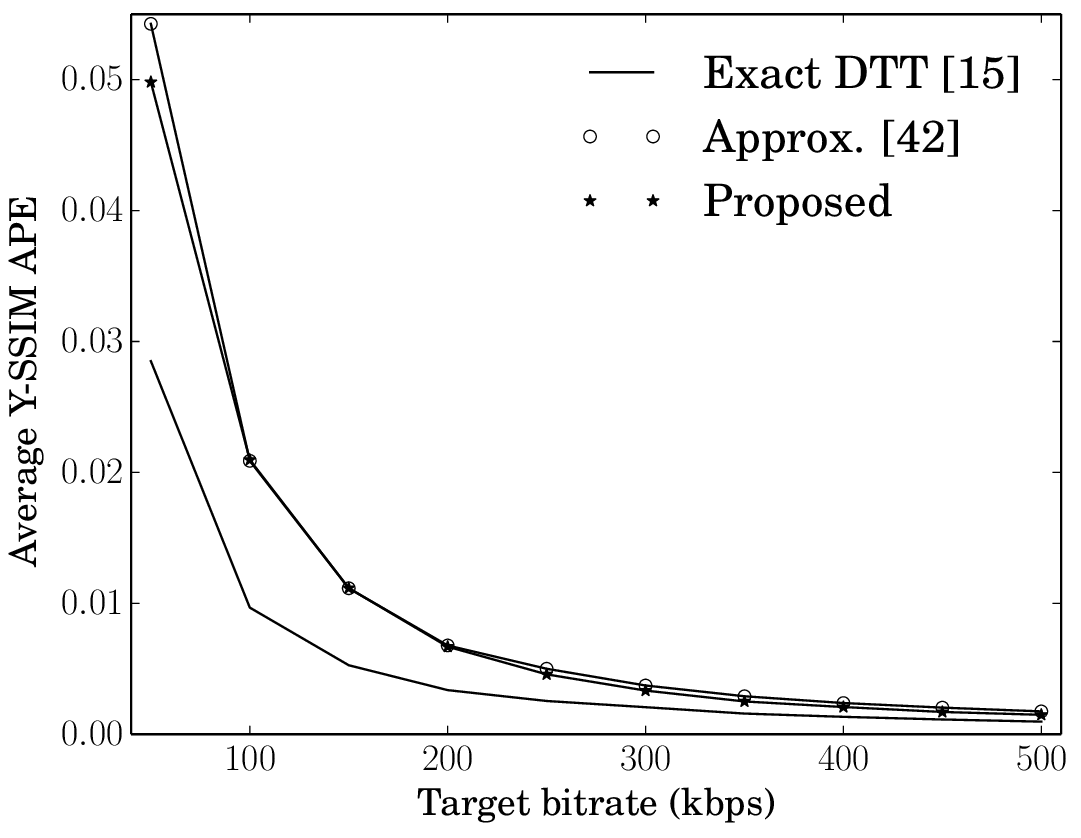}}
\subfigure[]{\includegraphics[width=0.4\textwidth]{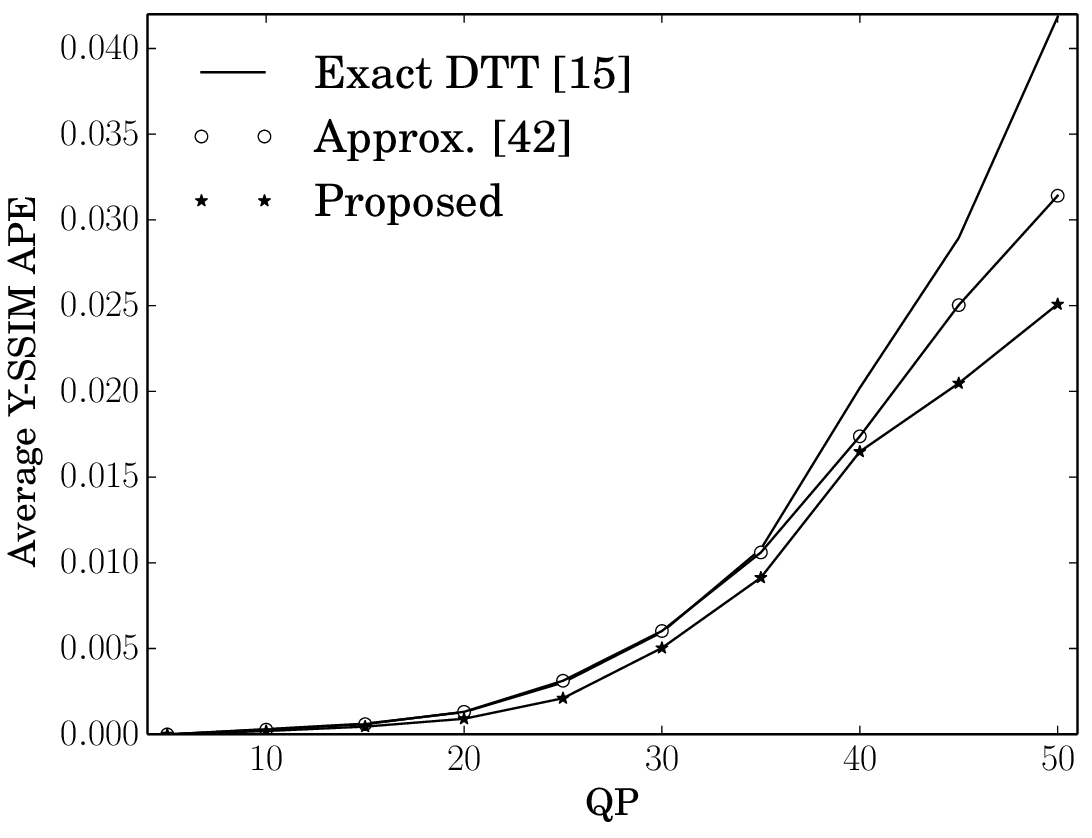}}
\caption{Video quality assessment in terms of target bitrate and~QP.}
\label{figure:bitrate}
\end{figure*}

Eleven
300-frame
common intermediate format~(CIF)
videos
obtained
from an online test video database~\cite{videos}
were encoded
with the standard software
and
then
with the modified software.
We employed the software default settings
and conducted the simulation
under two scenarios:
(i)~target bitrate varying from 50 to 500~kbps with steps of 50~kbps
and
(ii)~quantization parameter (QP) varying from 5 to 50 with steps of~5.
Psychovisual optimization was disabled
in order to obtain valid SSIM values.
Besides PSNR evaluation,
the discussed software library~\cite{x264} offers natively
SSIM measurements
for
video quality assessment.
Average SSIM of the luma component
were computed for all reconstructed frames.
The results are shown in Figure~\ref{figure:bitrate}
in terms of the absolute percentage error (APE)~\cite{cintra2014dct_aprox}
of the SSIM
with respect to the standard DCT-based transformation
in the original \mbox{H.264}/AVC codec.
This measure is given by:
\begin{align*}
\operatorname{APE}
(\mathrm{SSIM})
=
\left|
\frac
{\mathrm{SSIM}_\text{H.264} - \mathrm{SSIM}_{\mathbf{P}}}
{\mathrm{SSIM}_\text{H.264}}
\right|
,
\end{align*}
where
$\mathrm{SSIM}_\text{H.264}$
returns the SSIM figures
as computed according to
the \mbox{H.264} standard
and
$\mathrm{SSIM}_{\mathbf{P}}$
represents the SSIM
when
the exact DTT,
the approximation in~\cite{Oliveira2015Tchebichef},
or
the proposed approximation
are considered.
SSIM curves for the DCT are absent,
because they were employed as performance references.
The use of the proposed transform
effects a minor degradation in the video quality.
It also could perform better
than the previous approximation
in all cases.

\begin{figure*}
\centering
\subfigure[Foreman, H.264]{\includegraphics[width=0.25\textwidth]{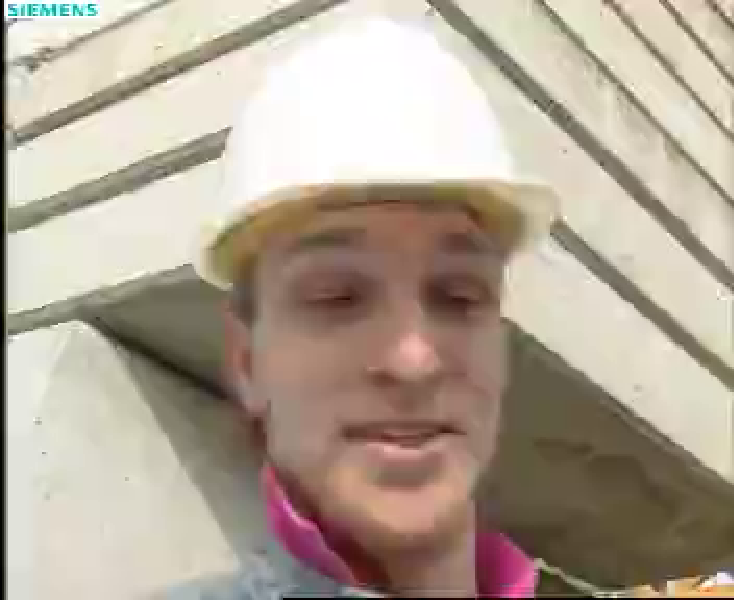}}
\qquad
\subfigure[Foreman, Approx.~\cite{Oliveira2015Tchebichef}]{\includegraphics[width=0.25\textwidth]{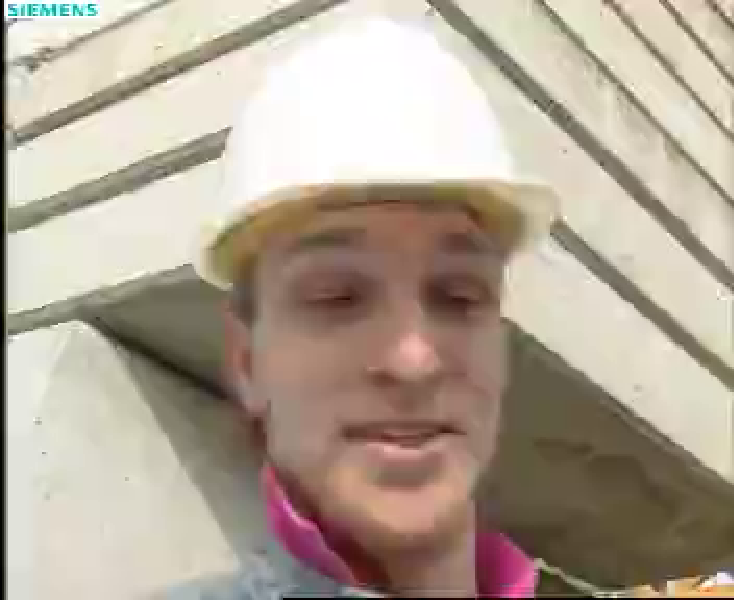}}
\qquad
\subfigure[Foreman, Proposed]{\includegraphics[width=0.25\textwidth]{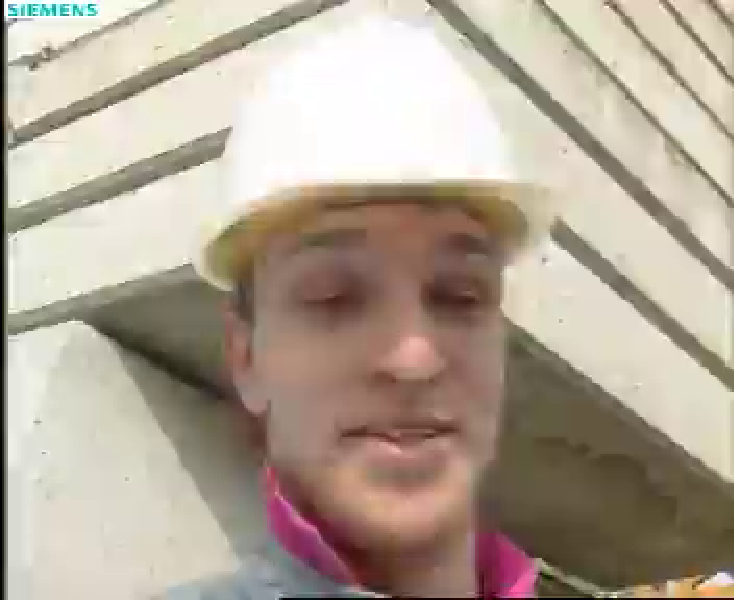}}
\caption{First frame of the compressed `Foreman' sequence, with a target bitrate of 200 kbps.}
\label{figure:video}
\end{figure*}

Figure~\ref{figure:video}
displays
the first encoded frame of two standard
video sequences at low bitrate (200 kbps).
The compressed frames resulting
from the original and modified codecs
are visually indistinguishable.

\section{Hardware Section}
\label{section-hardware}

To compare
the hardware resource consumption of
the proposed approximate DTT against
the exact DTT fast algorithm proposed in~\cite{Swamy2013ITT}, algorithm proposed in~\cite{Oliveira2015Tchebichef} and the Loeffler DCT~\cite{Loeffler1989},
the \mbox{2-D} version of both algorithms were
initially modeled and tested
in Matlab Simulink and
then were physically realized on
a Xilinx Virtex-6 XC6VSX475T-1FF1759 Reconfigurable Open Architecture Computing Hardware-2 (ROACH2) board~\cite{roach}.
The ROACH2 board consists of
a Xilinx Virtex~6 FPGA,
16~complex analog-to-digital converters (ADC),
multi-gigabit transceivers and a 72-bit DDR3 RAM.

The \mbox{1-D} versions were initially modeled and the \mbox{2-D} versions were generated using two \mbox{1-D} designs along with a transpose buffer.
Designs were verified using more than 10000~test vectors
with complete agreement with theoretical values.
Results are shown
in Table~\ref{FPGAresults}.
Metrics,
including
configurable logic blocks (CLB) and flip-flop (FF)
count,
critical path delay ($T_\text{cpd}$, in~ns), and
maximum operating frequency ($F_\text{max}$, in~MHz)
are provided.
The percentage reduction in the number of CLBs and FFs
were 43.2\% and 25.0\%,
respectively, compared with
the exact DTT fast algorithm proposed in~\cite{Swamy2013ITT}.
In is important to emphasize
that
the approximation in~\cite{Oliveira2015Tchebichef}
is asymmetric;
the forward and inverse transform possess
different structures, being the inverse operation
more complex
(cf.~Table~\ref{table:performance_compl}).
For comparisons,
we adopt the average measurement
between forward and inverse realizations.
The proposed approximation could
provide higher
maximum operating frequency
with improvements
of
85.9\%,
43.5\%,
and
9.7\%
when compared to
the Loeffler DCT~\cite{Loeffler1989},
the exact DTT~\cite{Swamy2013ITT},
and
the design in~\cite{Oliveira2015Tchebichef},
respectively.

\begin{table*}
\centering

\caption{Hardware resource consumption using Xilinx Virtex-6 XC6VSX475T 1FF1759 device}
\label{FPGAresults}
\begin{tabular}{lcccccc} %
\toprule
Method &
CLB &
FF &
$T_\text{cpd}$ \!\!\! ($\mathrm{ns}$) &
$F_{\text{max}}$ ($\mathrm{MHz}$)
\\
\midrule

Exact DTT~\cite{Swamy2013ITT} & 2941 & 7271 & 7.688 & 130.07
\\

\midrule

Approximation in~\cite{Oliveira2015Tchebichef} & 1515 & 6058 & 5.596 & 178.69
\\

\midrule

Inverse Approximation in~\cite{Oliveira2015Tchebichef} & 1713 & 4834 & 6.184 & 161.71
\\

\midrule

Loeffler DCT~\cite{Loeffler1989} & 3250 & 4413 & 9.956 & 100.44
\\

\midrule
Proposed DTT & 1671 & 5455 & 5.356 & 186.70
\\

\bottomrule
\end{tabular}

\end{table*}

The ASIC realization was done by porting the hardware description language code to 0.18~um CMOS technology and was subjected to synthesis and place-and-route
according to the Cadence Encounter Digital Implementation (EDI) for AMS libraries.
Libraries for the best case scenario were employed
in getting the place-and-route results with gate voltage of 1.8~V.
The adopted figures of merit for the ASIC synthesis
were:
area ($A$) in~$\mathrm{mm^2}$,
area-time complexity ($AT$) in $\mathrm{mm}^2 \cdot \mathrm{ns}$,
area-time-squared complexity ($AT^2$) in $\mathrm{mm}^2 \cdot \mathrm{ns}^2$,
dynamic ($D_p$) power consumption in mW/MHz,
critical path delay ($T_{cpd}$) in~ns,
and
maximum operating frequency ($F_{\text{max}}$) in MHz.
Results are displayed in Table~\ref{ASICResults}.
The figures of merit
$AT$ and $AT^2$ had percentage reductions of 57.7\% and 57.4\%
when compared with the exact DTT.
Thus,
the proposed design could attain
reductions of
17.3\%,
20.6\%,
and
82.1\%
for
area,
$AT^2$,
$T_\text{cpd}$,
and
dynamic power consumption,
respectively,
when compared
to~\cite{Oliveira2015Tchebichef}.

\begin{table*}
\centering

\caption{Hardware resource consumption for CMOS 0.18~um ASIC place and route}
\label{ASICResults}
\begin{tabular}{l@{\quad}c@{\quad}c@{\quad}c@{\quad}c@{\quad}c@{\quad}c@{\quad}c} %
\toprule
\parbox{2cm}{\centering FRS Method} &
 \parbox{2cm}{\centering Area ($\mathrm{mm}^2$)} &
\parbox{0.5cm}{\centering $AT$ } &
$AT^2$  &
$T_\text{cpd}$ ($\mathrm{ns}$) &
$F_{\text{max}}$ ($\mathrm{MHz}$) &
$D_p$ ($\mathrm{mW/MHz}$) \\ \midrule

Exact DTT\cite{Swamy2013ITT} &
0.872 & 3.84 & 16.92 & 4.405 & 227.01 & 0.182
\\
\midrule
Approximation in~\cite{Oliveira2015Tchebichef} &
0.237 & 1.34 & 7.52 & 5.635 & 177.46 & 0.171
\\
\midrule
Inverse Approximation in~\cite{Oliveira2015Tchebichef} &
0.323 & 1.79 & 9.89 & 5.536 & 180.64 & 0.724
\\
\midrule
Loeffler DCT~\cite{Loeffler1989} &
0.684 & 4.20 & 25.85 & 6.148 & 162.65 & 1.961
\\
\midrule
Proposed &
0.366 & 1.62 & 7.20 & 4.434 & 225.53 & 0.080
\\
\bottomrule

\end{tabular}
\end{table*}

\section{Discussion and Conclusion}
\label{section-conclusion}

In this work,
a
low-complexity
near-orthogonal
8-point DTT approximation
suitable
for
image and video coding
was proposed.
A fast algorithm for proposed DTT approximation
which requires only 24~additions and six bit-shifting operations
was also introduced.
This fast algorithm
can be used for both forward and near inverse transformations.
The additive arithmetic cost of the proposed approximation
is 45.5\%
and
2.05\%
lower
when compared with the
exact DTT fast algorithm
and
the DTT approximation~in~\cite{Oliveira2015Tchebichef},
respectively.
Moreover,
the proposed transform
exhibited similar coding performance with the exact DTT
and outperformed previous approximations~\cite{Oliveira2015Tchebichef}
according
to computational experiments
with popular visual image compression standards.
In terms of video coding,
the results from the proposed tool were
virtually indistinguishable from the ones
furnished by the approximation in~\cite{Oliveira2015Tchebichef}.
Thus,
the new tool outperform
the competing methods
both in computational cost and coding performance.
The proposed method was embedded into
the JPEG standard and
the standard software library for \mbox{H.264}/AVC video coding.
Obtained results showed negligible
degradation when compared to
the standard DCT-based
compression methods in both cases.
The \mbox{2-D} versions were realized in FPGA using ROACH2 hardware platform and ASIC place and route
was realized  using Cadence encounter with AMS standard cells and
the results show a $43.1\%$ reduction in the number of CLB
for the FPGA realization and a $57.7\%$ reduction in
area-time figure for the ASIC place and route realization when compared with the exact DTT.
The proposed design
could excel in providing high operation frequency
and
very low power consumption.
Therefore,
the proposed approximation
offers low computational complexity
while maintaining
good coding performance.
Systems that operate under
low processing constraints
and require video streaming
can benefit of the proposed
low-complexity
codecs and
low-power hardware.
In particular,
applications
in the following contexts
meet such requirements
that need low-complexity~\cite{WMSN2007survey}:
environmental monitoring, habitat
monitoring, surveillance, structural monitoring,
equipment diagnostics, disaster management, and
emergency response~\cite{Kimura2005}.

\section*{Acknowledgment}

Arjuna Madanayake thanks
the Xilinx University Program (XUP)
for the Xilinx Virtex-6 Sx475 FPGA device
installed in on the ROACH2 board.

{\footnotesize
\bibliographystyle{IEEEtran}
\bibliography{adtt_ref}

\begin{thebibliography}{10}
\providecommand{\url}[1]{#1}
\csname url@samestyle\endcsname
\providecommand{\newblock}{\relax}
\providecommand{\bibinfo}[2]{#2}
\providecommand{\BIBentrySTDinterwordspacing}{\spaceskip=0pt\relax}
\providecommand{\BIBentryALTinterwordstretchfactor}{4}
\providecommand{\BIBentryALTinterwordspacing}{\spaceskip=\fontdimen2\font plus
\BIBentryALTinterwordstretchfactor\fontdimen3\font minus
  \fontdimen4\font\relax}
\providecommand{\BIBforeignlanguage}[2]{{%
\expandafter\ifx\csname l@#1\endcsname\relax
\typeout{** WARNING: IEEEtran.bst: No hyphenation pattern has been}%
\typeout{** loaded for the language `#1'. Using the pattern for}%
\typeout{** the default language instead.}%
\else
\language=\csname l@#1\endcsname
\fi
#2}}
\providecommand{\BIBdecl}{\relax}
\BIBdecl

\bibitem{Nikiforov1991poly_discrete}
A.~F. Nikiforov, S.~K. Suslov, and V.~B. Uvarov, \emph{Classical Orthogonal
  Polynomials of a Discrete Variable}, ser. Springer Series in Computational
  Physics.\hskip 1em plus 0.5em minus 0.4em\relax Springer Berlin Heidelberg,
  1991.

\bibitem{Dragnev1997func_analysis}
\BIBentryALTinterwordspacing
P.~D. Dragnev and E.~B. Saff, ``\BIBforeignlanguage{English}{Constrained energy
  problems with applications to orthogonal polynomials of a discrete
  variable},'' \emph{\BIBforeignlanguage{English}{Journal d'Analyse
  Mathematique}}, vol.~72, pp. 223--259, 1997. [Online]. Available:
  \url{http://dx.doi.org/10.1007/BF02843160}
\BIBentrySTDinterwordspacing

\bibitem{Camara2009graphs}
M.~C\^{a}mara, J.~F\'{a}brega, M.~A. Fiol, and E.~Garriga, ``Some families of
  orthogonal polynomials of a discrete variable and their applications to
  graphs and codes,'' \emph{The Electronic Journal of Combinatorics}, vol.~16,
  pp. 1--30, 2009.

\bibitem{Zhu2010orth_moments}
H.~Zhu, M.~Liu, H.~Shu, H.~Zhang, and L.~Luo, ``General form for obtaining
  discrete orthogonal moments,'' \emph{IET Image Processing}, vol.~4, pp.
  335--352, Oct 2010.

\bibitem{Goshtasby1985moment1}
A.~Goshtasby, ``Template matching in rotated images,'' \emph{IEEE Transactions
  on Pattern Analysis and Machine Intelligence}, vol.~7, no.~3, pp. 338--344,
  May 1985.

\bibitem{Heywood1995moment2}
M.~I. Heywood and P.~D. Noakes, ``Fractional central moment method for
  movement-invariant object classification,'' \emph{IEE Proceedings--Vision,
  Image and Signal Processing}, vol. 142, no.~4, pp. 213--219, Aug 1995.

\bibitem{Markandey1992moment3}
V.~Markandey and R.~I.~P. de~Figueiredo, ``Robot sensing techniques based on
  high-dimensional moment invariants and tensors,'' \emph{IEEE Transactions on
  Robotics and Automation}, vol.~8, no.~2, pp. 186--195, Apr 1992.

\bibitem{DTM}
R.~Mukundan, S.~H. Ong, and R.~A. Lee, ``Image analysis by {T}chebichef
  moments,'' \emph{IEEE Transactions on Image Processing}, vol.~10, pp.
  1357--1364, 2001.

\bibitem{Leida2014artifact_tchebichef}
L.~Leida, Z.~Hancheng, Y.~Gaobo, and Q.~Jiansheng, ``Referenceless measure of
  blocking artifacts by {T}chebichef kernel analysis,'' \emph{IEEE Signal
  Processing Letters}, vol.~21, no.~1, pp. 122--125, Jan 2014.

\bibitem{Rose2009shape_tchebichef}
J.-L. Rose, C.~Revol-Muller, D.~Charpigny, and C.~Odet, ``Shape prior criterion
  based on {T}chebichef moments in variational region growing,'' in \emph{2009
  16th IEEE International Conference on Image Processing (ICIP)}, Nov 2009, pp.
  1081--1084.

\bibitem{Zhang2010img_recog}
H.~Zhang, X.~Dai, P.~Sun, H.~Zhu, and H.~Shu, ``Symmetric image recognition by
  {T}chebichef moment invariants,'' in \emph{2010 17th IEEE International
  Conference on Image Processing (ICIP)}, Sept 2010, pp. 2273--2276.

\bibitem{Li2011recognition_tchebichef}
Q.~Li, H.~Zhu, and Q.~Liu, ``Image recognition by combined affine and blur
  {T}chebichef moment invariants,'' in \emph{2011 4th International Congress on
  Image and Signal Processing (CISP)}, vol.~3, Oct 2011, pp. 1517--1521.

\bibitem{Roux2012blind}
H.~Huang, G.~Coatrieux, H.~Shu, L.~Luo, and C.~Roux, ``Blind integrity
  verification of medical images,'' \emph{IEEE Transactions on Information
  Technology in Biomedicine}, vol.~16, no.~6, pp. 1122--1126, Nov 2012.

\bibitem{Swamy2008DTT}
S.~Ishwar, P.~K. Meher, and M.~N.~S. Swamy, ``Discrete {T}chebichef
  transform--a fast 4 $\times$ 4 algorithm and its application in image/video
  compression,'' in \emph{2008 IEEE International Symposium on Circuits and
  Systems (ISCAS)}, 2008, pp. 260--263.

\bibitem{Swamy2013ITT}
S.~Prattipati, S.~Ishwar, P.~K. Meher, and M.~N.~S. Swamy, ``A fast 8$\times$8
  integer {T}chebichef transform and comparison with integer cosine transform
  for image compression,'' in \emph{2013 IEEE 56th International Midwest
  Symposium on Circuits and Systems (MWSCAS)}, 2013, pp. 1294--1297.

\bibitem{Mukundan2010img_compress}
N.~A. Abu, S.~L. Wong, N.~Herman, and R.~Mukundan, ``An efficient compact
  {T}chebichef moment for image compression,'' in \emph{2010 10th International
  Conference on Information Sciences Signal Processing and their Applications
  (ISSPA)}, May 2010, pp. 448--451.

\bibitem{Li2012cs_tchebichef}
Q.~Li and H.~Zhu, ``Block-based compressed sensing of image using directional
  {T}chebichef transforms,'' in \emph{2012 IEEE International Conference on
  Systems, Man, and Cybernetics (SMC)}, Oct 2012, pp. 2207--2212.

\bibitem{Senapati2014listlessDTT}
R.~K. Senapati, U.~C. Pati, and K.~K. Mahapatra, ``Reduced memory, low
  complexity embedded image compression algorithm using hierarchical listless
  discrete {T}chebichef transform,'' \emph{IET Image Processing}, vol.~8,
  no.~4, pp. 213--238, Apr 2014.

\bibitem{Ahmed1974DCT}
N.~Ahmed, T.~Natarajan, and K.~R. Rao, ``Discrete cosine transform,''
  \emph{IEEE Transactions on Computers}, vol. C-23, no.~1, pp. 90--93, Jan.
  1974.

\bibitem{Ernawan2013quantization_tchebichef}
F.~Ernawan, N.~A. Abu, and N.~Suryana, ``{TMT} quantization table generation
  based on psychovisual threshold for image compression,'' in \emph{2013
  International Conference of Information and Communication Technology
  (ICoICT)}, Mar 2013, pp. 202--207.

\bibitem{Senapati2011DTT_coding}
R.~K. Senapati, U.~C. Pati, and K.~K. Mahapatra, ``A low complexity embedded
  image coding algorithm using hierarchical listless {DTT},'' in \emph{2011 8th
  International Conference on Information, Communications and Signal Processing
  (ICICS)}, Dec 2011, pp. 1--5.

\bibitem{Ernawan2011mobile_tchebichef}
F.~Ernawan, E.~Noersasongko, and N.~A. Abu, ``An efficient 2$\times$2
  {T}chebichef moments for mobile image compression,'' in \emph{2011
  International Symposium on Intelligent Signal Processing and Communications
  Systems (ISPACS)}, Dec 2011, pp. 1--5.

\bibitem{Li2008WSN}
L.~W. Chew, L.-M. Ang, and K.~P. Seng, ``Survey of image compression algorithms
  in wireless sensor networks,'' in \emph{2008 International Symposium on
  Information Technology (ITSim)}, vol.~4, Aug 2008, pp. 1--9.

\bibitem{Meng2005realtime_video}
M.~Guo, M.~H. Ammar, and E.~W. Zegura, ``{V}3: a vehicle-to-vehicle live video
  streaming architecture,'' in \emph{2005 3rd IEEE International Conference on
  Pervasive Computing and Communication (PerCom)}, Mar 2005, pp. 171--180.

\bibitem{Friedman2013video_low_power}
D.~H. Friedman, ``Streaming implementation of video algorithms on a low-power
  parallel architecture,'' in \emph{2013 IEEE Global Conference on Signal and
  Information Processing (GlobalSIP)}, Dec 2013, pp. 650--653.

\bibitem{Mukundan2007FDTT}
K.~Nakagaki and R.~Mukundan, ``A fast 4$\times$4 forward discrete {T}chebichef
  transform algorithm,'' \emph{IEEE Signal Processing Letters}, vol.~14, pp.
  684--687, 2007.

\bibitem{Wallace1992JPEG}
G.~K. Wallace, ``The {JPEG} still picture compression standard,'' \emph{IEEE
  Transactions on Consumer Electronics}, vol.~38, no.~1, pp. xviii--xxxiv, Feb
  1992.

\bibitem{MPEG-2}
{International Organisation for Standardisation}, ``Generic coding of moving
  pictures and associated audio information – part 2: Video, {ISO/IEC}
  {JTC1/SC29/WG11} – coding of moving pictures and audio,'' 1994.

\bibitem{H.261}
{International Telecommunication Union}, ``{ITU-T} recommendation {H}.261
  version 1: Video codec for audiovisual services at $p \times 64$ kbits,''
  Technical Report, {ITU-T}, 1990.

\bibitem{H.263}
------, ``{ITU-T} recommendation {H}.263 version 1: Video coding for low bit
  rate communication,'' Technical Report, {ITU-T}, 1995.

\bibitem{H264_book}
I.~Richardson, \emph{The {H}.264 Advanced Video Compression Standard},
  2nd~ed.\hskip 1em plus 0.5em minus 0.4em\relax John Wiley and Sons, 2010.

\bibitem{Sullivan2012HEVC}
G.~J. Sullivan, J.~Ohm, W.-J. Han, and T.~Wiegand, ``Overview of the high
  efficiency video coding ({HEVC}) standard,'' \emph{IEEE Transactions on
  Circuits and Systems for Video Technology}, vol.~22, pp. 1649--1668, 2012.

\bibitem{Bossen2012HEVC_impementation}
F.~Bossen, B.~Bross, K.~Suhring, and D.~Flynn, ``{HEVC} complexity and
  implementation analysis,'' \emph{IEEE Transactions on Circuits and Systems
  for Video Technology}, vol.~22, no.~12, pp. 1685--1696, Dec 2012.

\bibitem{VP9}
{Google Inc.}, ``{VP9},'' The WebM Project,
  \url{http://www.webmproject.org/vp9/}, 2015.

\bibitem{Haweel2001SDCT}
\BIBentryALTinterwordspacing
T.~I. Haweel, ``A new square wave transform based on the {DCT},'' \emph{Signal
  Processing}, vol.~81, pp. 2309--2319, 2001. [Online]. Available:
  \url{http://www.sciencedirect.com/science/article/pii/S0165168401001062}
\BIBentrySTDinterwordspacing

\bibitem{CB2011RDCT}
R.~J. Cintra and F.~M. Bayer, ``A {DCT} approximation for image compression,''
  \emph{IEEE Signal Processing Letters}, vol.~18, no.~10, pp. 579--582, Oct
  2011.

\bibitem{CB2012MRDCT}
F.~M. Bayer and R.~J. Cintra, ``{DCT}-like transform for image compression
  requires 14 additions only,'' \emph{Electronics Letters}, vol.~48, no.~15,
  pp. 919--921, July 2012.

\bibitem{cintra2014dct_aprox}
\BIBentryALTinterwordspacing
R.~J. Cintra, F.~M. Bayer, and C.~J. Tablada, ``Low-complexity 8-point {DCT}
  approximations based on integer functions,'' \emph{Signal Processing},
  vol.~99, pp. 201--214, 2014. [Online]. Available:
  \url{http://www.sciencedirect.com/science/article/pii/S0165168413005161}
\BIBentrySTDinterwordspacing

\bibitem{BAS2008_EL}
S.~Bouguezel, M.~O. Ahmad, and M.~N.~S. Swamy, ``Low-complexity 8$\times$8
  transform for image compression,'' \emph{Electronics Letters}, vol.~44,
  no.~21, pp. 1249--1250, Oct 2008.

\bibitem{BAS2011}
------, ``A low-complexity parametric transform for image compression,'' in
  \emph{2011 IEEE International Symposium on Circuits and Systems (ISCAS)}, May
  2011, pp. 2145--2148.

\bibitem{BAS2013}
------, ``Binary discrete cosine and hartley transforms,'' \emph{IEEE
  Transactions on Circuits and Systems I: Regular Papers}, vol.~60, no.~4, pp.
  989--1002, Apr 2013.

\bibitem{Oliveira2015Tchebichef}
P.~A.~M. Oliveira, R.~J. Cintra, F.~M. Bayer, S.~Kulasekera, and A.~Madanayake,
  ``A discrete {T}chebichef transform approximation for image and video
  coding,'' \emph{IEEE Signal Processing Letters}, vol.~22, no.~8, pp.
  1137--1141, Aug 2015.

\bibitem{HTF53}
\BIBentryALTinterwordspacing
H.~Bateman and A.~Erd{\'e}lyi, \emph{Higher transcendental functions}.\hskip
  1em plus 0.5em minus 0.4em\relax McGraw-Hill, 1953, vol.~2. [Online].
  Available: \url{http://books.google.com.br/books?id=p\_lQAAAAMAAJ}
\BIBentrySTDinterwordspacing

\bibitem{malvar2003H264matrix}
H.~S. Malvar, A.~Hallapuro, M.~Karczewicz, and L.~Kerofsky, ``Low-complexity
  transform and quantization in {H.264/AVC},'' \emph{IEEE Transactions on
  Circuits and Systems for Video Technology}, vol.~13, no.~7, pp. 598--603, Jul
  2003.

\bibitem{blahut_book}
R.~Blahut, \emph{Fast Algorithms for Signal Processing}.\hskip 1em plus 0.5em
  minus 0.4em\relax Cambridge University Press, 2010.

\bibitem{MATLAB}
MATLAB, ``version 8.1 ({R}2013a) documentation,'' Natick, MA, 2013.

\bibitem{octave}
J.~W. Eaton, D.~Bateman, S.~Hauberg, and R.~Wehbring, \emph{GNU Octave version
  3.8.0 Documentation}, 3rd~ed.\hskip 1em plus 0.5em minus 0.4em\relax Free
  Software Foundation, Inc., Feb 2011.

\bibitem{python}
Python, ``version 2.7.6 documentation,'' Delaware, US, 2015.

\bibitem{seber_matrix}
G.~A.~F. Seber, \emph{A Matrix Handbook for Statisticians}, ser. Wiley Series
  in Probability and Mathematical Statistics.\hskip 1em plus 0.5em minus
  0.4em\relax Hoboken, NJ: John Wiley and Sons, Inc., 2008.

\bibitem{flury86}
\BIBentryALTinterwordspacing
B.~N. Flury and W.~Gautschi, ``An algorithm for simultaneous orthogonal
  transformation of several positive definite symmetric matrices to nearly
  diagonal form,'' \emph{SIAM Journal on Scientific and Statistical Computing},
  vol.~7, no.~1, pp. 169--184, Jan. 1986. [Online]. Available:
  \url{http://dx.doi.org/10.1137/0907013}
\BIBentrySTDinterwordspacing

\bibitem{Goyal2001coding}
V.~K. Goyal, ``Theoretical foundations of transform coding,'' \emph{IEEE Signal
  Processing Magazine}, vol.~18, no.~5, pp. 9--21, Sept 2001.

\bibitem{Katto1991cg}
\BIBentryALTinterwordspacing
J.~Katto and Y.~Yasuda, ``Performance evaluation of subband coding and
  optimization of its filter coefficients,'' \emph{Journal of Visual
  Communication and Image Representation}, vol.~2, pp. 303--313, 1991.
  [Online]. Available:
  \url{http://www.sciencedirect.com/science/article/pii/1047320391900114}
\BIBentrySTDinterwordspacing

\bibitem{britanak_book}
\BIBentryALTinterwordspacing
V.~Britanak, P.~C. Yip, and K.~R. Rao, \emph{Discrete Cosine and Sine
  Transforms}.\hskip 1em plus 0.5em minus 0.4em\relax Academic Press, 2007.
  [Online]. Available: \url{http://books.google.com.br/books?id=iRlQHcK-r\_kC}
\BIBentrySTDinterwordspacing

\bibitem{tablada2015feig_winograd}
C.~J. Tablada, F.~M. Bayer, and R.~J. Cintra, ``A class of {DCT} approximations
  based on the {Feig--Winograd} algorithm,'' \emph{Signal Processing}, vol.
  113, pp. 38--51, 2015.

\bibitem{Cintra2002RHT}
R.~J. Cintra, H.~M. Oliveira, and C.~O. Cintra, ``The rounded {H}artley
  transform,'' in \emph{Proceedings of the IEEE International
  Telecommunications Symposium--ITS'2002}, Sept 2002, pp. 1357--1364.

\bibitem{Pennebaker1993JPEG}
W.~B. Pennebaker and J.~L. Mitchell, \emph{JPEG: Still Image Data Compression
  Standard}, ser. Chapman \& Hall digital multimedia standards series.\hskip
  1em plus 0.5em minus 0.4em\relax Springer, 1993.

\bibitem{bovik2009mse}
Z.~Wang and A.~C. Bovik, ``Mean squared error: Love it or leave it? a new look
  at signal fidelity measures,'' \emph{IEEE Signal Processing Magazine},
  vol.~26, no.~1, pp. 98--117, Jan 2009.

\bibitem{Chi2012distortion}
C.-K. Fong and W.-K. Cham, ``{LLM} integer cosine transform and its fast
  algorithm,'' \emph{IEEE Transactions on Circuits and Systems for Video
  Technology}, vol.~22, no.~6, pp. 844--854, Jun 2012.

\bibitem{Bayer2013multless4point}
F.~M. Bayer, R.~J. Cintra, A.~Madanayake, and U.~S. Potluri, ``Multiplierless
  approximate 4-point {DCT} {VLSI} architectures for transform block coding,''
  \emph{Electronics Letters}, vol.~49, no.~24, pp. 1532--1534, Nov 2013.

\bibitem{oppenheim_book}
A.~V. Oppenheim and R.~W. Schafer, \emph{Discrete-time signal processing},
  3rd~ed., ser. Prentice-Hall signal processing series.\hskip 1em plus 0.5em
  minus 0.4em\relax Prentice Hall, 2010.

\bibitem{Loeffler1989}
C.~Loeffler, A.~Ligtenberg, and G.~S. Moschytz, ``A practical fast 1-{D} {DCT}
  algorithms with 11 multiplications,'' in \emph{IEEE International Conference
  on Acoustics, Speech, and Signal Processing}, vol.~2, May 1989, pp. 988--991.

\bibitem{imagens}
{University of Southern California, Signal and Image Processing Institute},
  ``The {USC-SIPI} image database,'' \url{http://sipi.usc.edu/database/}, 2015.

\bibitem{Bovik2004SSIM}
Z.~Wang, A.~C. Bovik, H.~R. Sheikh, and E.~P. Simoncelli, ``Image quality
  assessment: from error visibility to structural similarity,'' \emph{IEEE
  Transactions on Image Processing}, vol.~13, no.~4, pp. 600--612, Apr 2004.

\bibitem{zhang2012SR_SIM}
L.~Zhang and H.~Li, ``{SR-SIM}: A fast and high performance {IQA} index based
  on spectral residual,'' in \emph{2012 19th IEEE International Conference on
  Image Processing (ICIP)}, Sep 2012, pp. 1473--1476.

\bibitem{Bovik2011SSIM}
Z.~Wang and A.~C. Bovik, ``Reduced- and no-reference image quality
  assessment,'' \emph{IEEE Signal Processing Magazine}, vol.~28, no.~6, pp.
  29--40, Nov 2011.

\bibitem{kay_book}
S.~M. Kay, \emph{Fundamentals of Statistical Signal Processing, Volume I:
  Estimation Theory}, ser. Prentice Hall Signal Processing Series.\hskip 1em
  plus 0.5em minus 0.4em\relax Upper Saddle River, NJ: Prentice-Hall, 1993,
  vol.~1.

\bibitem{Pandit2013quality}
R.~Pandit, N.~Khosla, G.~Singh, , and H.~Sharma, ``Image compression and
  quality factor in case of {JPEG} image format,'' \emph{International Journal
  of Advanced Research in Computer and Communication Engineering}, vol.~2, pp.
  2578--2581, Jul 2013.

\bibitem{x264}
x264 team, ``x264,'' \url{http://www.videolan.org/developers/x264.html}, 2015.

\bibitem{H264_8transform}
S.~Gordon, D.~Marpe, and T.~Wiegand, ``Simpliﬁed use of 8$\times$8
  transform--updated proposal and results,'' Joint Video Team ({JVT}) of
  {ISO/IEC MPEG} and {ITU-T VCEG}, doc. {JVT}--{K028}, Munich, Germany, Mar
  2004.

\bibitem{videos}
``{Xiph.org Video Test Media},'' \url{https://media.xiph.org/video/derf/},
  2015.

\bibitem{roach}
(2015) {ROACH2}. \url{https://casper.berkeley.edu}.

\bibitem{WMSN2007survey}
I.~F. Akyildiz, T.~Melodia, and K.~R. Chowdhury, ``A survey on wireless
  multimedia sensor networks,'' \emph{Computer Networks}, vol.~51, pp.
  921--960, 2007.

\bibitem{Kimura2005}
N.~Kimura and S.~Latifi, ``A survey on data compression in wireless sensor
  networks,'' in \emph{2005 International Conference on Information Technology:
  Coding and Computing (ITCC)}, vol.~2, Apr 2005, pp. 8--13.

\end{thebibliography}
}

\end{document}